\documentclass{article}

\usepackage{graphicx}
\usepackage{microtype}

\usepackage{lineno}
\usepackage{amsmath,amsfonts,amssymb}
\usepackage{ntheorem}

\theoremstyle{plain}
\theorembodyfont{\normalfont}

\theoremstyle{plain}
\theorembodyfont{\normalfont}

\usepackage{etoolbox} 
\usepackage{siunitx}

\newcommand*\linenomathpatch[1]{%
  \cspreto{#1}{\linenomath}%
  \cspreto{#1*}{\linenomath}%
  \csappto{end#1}{\endlinenomath}%
  \csappto{end#1*}{\endlinenomath}%
}

\linenomathpatch{equation}
\linenomathpatch{gather}
\linenomathpatch{multline}
\linenomathpatch{align}
\linenomathpatch{alignat}
\linenomathpatch{flalign}

\usepackage{bm}
\usepackage[position=top,labelfont=normalfont,textfont=normalfont,singlelinecheck=off,justification=raggedright]{subcaption}
\usepackage{floatrow}
\usepackage[dvipsnames]{xcolor}
\usepackage{setspace}
\usepackage{enumerate,etaremune}
\usepackage{booktabs}
\usepackage{tabularx,multirow}
\usepackage{framed}
\usepackage{hhline}
\usepackage{url}
\usepackage[unicode,bookmarks=false]{hyperref}
\hypersetup{
  colorlinks,
  citecolor=Blue,linkcolor=Blue,urlcolor=Blue
}
\usepackage[toc,page]{appendix}
\usepackage[T1]{fontenc}

\usepackage{tikz}
\usetikzlibrary{shapes.geometric}
\usetikzlibrary{shapes,arrows}
\usetikzlibrary{plotmarks}
\usetikzlibrary{calc}
\usetikzlibrary{
  arrows,
  calc,
  fit,
  patterns,
  plotmarks,
  shapes.geometric,
  shapes.misc,
  shapes.symbols,
  shapes.arrows,
  shapes.callouts,
  shapes.multipart,
  shapes.gates.logic.US,
  shapes.gates.logic.IEC,
  er,
  automata,
  backgrounds,
  chains,
  topaths,
  trees,
  petri,
  mindmap,
  matrix,
  calendar,
  folding,
  fadings,
  through,
  positioning,
  scopes,
  decorations.fractals,
  decorations.shapes,
  decorations.text,
  decorations.pathmorphing,
  decorations.pathreplacing,
  decorations.footprints,
  decorations.markings,
  shadows}
\usetikzlibrary{tikzmark}

\usepackage{algorithm}
\usepackage{algorithmicx,algpseudocode}
\usepackage{cases}
\algdef{SE}[DOWHILE]{Do}{doWhile}{\algorithmicdo}[1]{\algorithmicwhile\ #1}

\def\eqref#1{(\ref{#1})}

\def\1{\bm{1}}

\def\vd{{\bm{d}}}

\def\vm{{\bm{m}}}

\def\vs{{\bm{s}}}

\def\vw{{\bm{w}}}
\def\vx{{\bm{x}}}
\def\vy{{\bm{y}}}

\DeclareMathAlphabet{\mathsfit}{\encodingdefault}{\sfdefault}{m}{sl}
\SetMathAlphabet{\mathsfit}{bold}{\encodingdefault}{\sfdefault}{bx}{n}

\def\sX{{\mathbb{X}}}

\newsavebox{\dotbox}

\newcolumntype{L}[1]{>{\raggedright\let\newline\\arraybackslash\hspace{0pt}}m{#1}}
\newcolumntype{C}[1]{>{\centering\let\newline\\arraybackslash\hspace{0pt}}m{#1}}
\newcolumntype{R}[1]{>{\raggedleft\let\newline\\arraybackslash\hspace{0pt}}m{#1}}

\allowdisplaybreaks

\definecolor{darkred}{rgb}{0.55, 0.0, 0.0}
\definecolor{skyblue}{rgb}{0.53, 0.81, 0.92}

\usepackage{hyperref}
\usepackage{soul}

\usepackage[numbers]{natbib}
\usepackage{arxiv}
\usepackage{fancyhdr}

\renewcommand{\headeright}{}
\renewcommand{\undertitle}{}
\usepackage{graphics}
\usepackage{authblk}
\usepackage{xcolor}
\usepackage{natbib}

\usepackage{authblk} 

\newcommand{\revision}[1]{\textcolor{black}{#1}} 
\newcommand{\asinh}[0]{\sinh^{-1}}
\newcommand{\ours}[0]{TGLF-WINN}
\newcommand{\oursfull}[0]{Wavenumber-Informed Neural Network}
\newcommand{\base}[0]{\text{TGLF-NN}}
\newcommand{\firstcontri}[0]{\text{FT}}
\newcommand{\secondcontri}[0]{\text{WR}}

\newcommand{\channelone}[0]{\Gamma_e}
\newcommand{\channeltwo}[0]{\Pi_i}
\newcommand{\channelthree}[0]{Q_e}
\newcommand{\channelfour}[0]{Q_i}
\newcommand{\MaPert}[0]{Major perturbation}
\newcommand{\MaMiPert}[0]{Major + Minor perturbation}

\begin{document}

\title{\ours: Data-Efficient Deep Learning Surrogate for Turbulent Transport Modeling in Fusion}

\author[1,*]{Yadi Cao}
\author[1,*]{Futian Zhang}
\author[1,*]{Wesley Liu}
\author[2]{Tom Neiser}
\author[2]{Orso Meneghini}
\author[1]{Lawson Fuller}
\author[2]{\\Sterling Smith}
\author[2]{Raffi Nazikian}
\author[2]{Brian Sammuli}
\author[1]{Rose Yu}
\affil[1]{University of California, San Diego, La Jolla, CA 92093, USA}
\affil[2]{General Atomics, P.O. Box 85608, San Diego, CA 92186, USA}
\affil[*]{Equal contribution}

\date{} 

\maketitle

\begin{abstract}
    The Trapped Gyro-Landau Fluid (TGLF) model provides fast, accurate predictions of turbulent transport in tokamaks, but whole device simulations requiring thousands of evaluations remain computationally expensive. Neural network (NN) surrogates offer accelerated inference with fully differentiable approximations that enable gradient-based coupling but typically require large training datasets to capture transport flux variations across plasma conditions, creating significant training burden and limiting applicability to expensive gyrokinetic simulations. We propose \textbf{\ours~(\oursfull)} with three key innovations: (1) principled feature engineering that reduces target prediction range, simplifying the learning task; (2) physics-guided wavenumber-resolved regularization to improve generalization under sparse data; and (3) Bayesian Active Learning (BAL) to strategically select training samples based on model uncertainty, reducing data requirements while maintaining accuracy. \ours~is engineered for data-efficient and robust surrogate training. Feature tuning and wavenumber regularization together deliver a 12.5\% relative RMSLE reduction over \base~when trained on the complete dataset; more importantly, under sparse, unfiltered training conditions (approximately $1/9$ the full dataset size) these two ingredients yield an order-of-magnitude smaller RMSLE degradation than \base, a robustness attributable to the wavenumber-informed regularization imposing a physics-guided constraint on per-mode flux contributions. Adding Bayesian Active Learning on top, \ours~matches \base's full-data offline accuracy using only 25\% of the training data, reaching RMSLE within 2.8\% of \base's full-data baseline and within 4.3\% of our own full-data result. We further demonstrate practicality in a downstream flux-matching workflow: the NN surrogate provides a 45$\times$ speedup over TGLF while maintaining comparable reconstruction accuracy.
\end{abstract}

\section{Introduction}


Comprehensive simulation of fusion devices (``whole-device'' modeling) is critical for understanding plasma behavior and optimizing reactor performance~\cite{bonoli2015report, suchyta2022exascale, Meneghini24}. Turbulent transport is central to these simulations, strongly coupled with neoclassical simulations~\cite{hirshman1981neoclassical,Belli12_NEO}, pedestal regions~\cite{snyder2009development,snyder2011first}, and equilibrium modeling~\cite{lao1996rotational, Lyons23, Slendebroek23, McClenaghan24}. Although gyrokinetic simulations~\cite{dimits2000simulation,candy2003eulerian,howard2015multi,candy2016high, Neiser19} provide the most comprehensive turbulent transport through five-dimensional phase space modeling, they are computationally expensive, requiring hours on supercomputers for single evaluations and making them impractical for coupled workflows.

Reduced order models (ROMs) based on quasi-linear theory~\cite{waltz1997gyro,candy2009tokamak,rafiq2013physics} mitigate these costs by performing linear instability analysis and then post-fixing with theory-based ``saturation rules''~\cite{staebler2007theory} that are calibrated on datasets generated by gyrokinetic simulations. The Trapped Gyro-Landau Fluid (TGLF) model~\cite{staebler2005gyro,staebler2007theory,kinsey2008first} represents the state of the art, which requires only seconds per evaluation versus hours for gyrokinetic simulations. However, whole-device simulations that require thousands of transport evaluations per time step remain computationally expensive.

Machine learning has accelerated physics simulations in all domains~\cite{pfaff2020learning,wu2021deepgleam,cao2023efficient,cao2024vicon}, with fusion applications in control optimization~\cite{char2019offline,seo2024avoiding} and surrogate modeling. Direct plasma profile evolution surrogates using CNNs and LSTMs~\cite{abbate2021data,Fenstermacher22} show significant prediction errors, while Fourier neural networks applied to MHD data~\cite{gopakumar2024plasma,rahman2024sparsified} use coarser modeling than kinetic theory. Moreover, directly surrogating plasma profile evolution is not practical, as it requires an exhaustive dataset that accounts for all factors from multiphysics components and control signals in tokamak operation.

In contrast, component-level surrogates for turbulent transport~\cite{citrin2015real,meneghini2017self,Neiser22,van2020fast,fransson2023fast} offer more practical alternatives. \base~\cite{meneghini2017self, Neiser22} trains neural networks to predict the TGLF output, reducing the evaluation time to microseconds. However, these methods require large datasets to capture wide parameter ranges and flux magnitude variations. QuaLiKiz-NN addresses this through mixture-of-experts (MOE) design~\cite{van2020fast,fransson2023fast}, dividing datasets by instability modes and promoting negative flux predictions during training while clipping to zero during inference. Yet, this approach has limitations: comprehensive models like TGLF don't inherently clip fluxes, and predictions remain sensitive to expert misclassification.

In this paper, we propose \textbf{\ours~(\oursfull)} to address data-efficiency challenges through three key improvements: 1) \textbf{Feature Tuning}: Systematic feature engineering and preprocessing with new loss terms that improve flux magnitude differentiability over \base; 2) \textbf{Wavenumber Regularization}: Incorporating per-wavenumber transport fluxes as regularization terms, leveraging TGLF's wavenumber integration structure; 3) \textbf{Bayesian Active Learning (BAL)}: Strategic data point selection using Expected Information Gain~\cite{settles2012active} to minimize training data requirements~\cite{pavone2023machine}.

Our contributions include: (1) \ours, incorporating principled feature engineering and wavenumber-informed regularization for equivalent accuracy to \base~(Section~\ref{sec:res:overall}); (2) demonstration of substantially improved robustness in noisy, sparse dataset settings (Section~\ref{sec:res:sparse}); (3) BAL integration reducing data requirements to 25\% of full datasets (Section~\ref{sec:res:bal}); (4) validation of our method in downstream flux-matching applications (Section~\ref{sec:res:fluxmatching}). Combined, our method shows a great potential for extension to higher-fidelity models like full gyrokinetic simulations~\cite{Neiser23} where data generation is prohibitively expensive.

\section{Related Work}
\paragraph{TGLF and Saturation Rules.}
TGLF~\cite{staebler2005gyro,staebler2007theory} is a quasi-linear transport model that projects gyrokinetic equations into moment equations. It then performs linear mode analysis at $N_y$ different wavenumbers \(\{k_{y1}^1, \dots, k_{y}^{N_y}\}\) around the thermal equilibrium distribution to obtain quasilinear weights and intensities, with nonlinear mode interactions post-modeled using empirical ``saturation rules''~\cite{waltz1999ion,kinsey2005predicting,staebler2007theory} that are calibrated on a dataset simulated by gyrokinetic simulations. TGLF is extensively validated and considered state-of-the-art for plasma studies~\cite{meneghini2017self, Neiser22}.

We use TGLF with the ``SAT2''~\cite{Staebler20, Staebler21, dudding2022new, Staebler24} saturation rule as our training targets.

\paragraph{Machine Learning Surrogates for Fusion.}
There are two distinct approaches for surrogate modeling in plasma: 1) learning complete plasma profile evolution (CNN+LSTM~\cite{abbate2021data}, FNO~\cite{gopakumar2024plasma,rahman2024sparsified}), and 2) learning components like core turbulent transport~\cite{citrin2015real,meneghini2017self, Neiser22,van2020fast,fransson2023fast}. The former suffers from high training difficulty and scalability issues due to high-dimensional inputs from multi-physics and control variables, making the curation datasets impractical. Component-level approaches narrow input ranges, enabling easier training and flexible downstream coupling. However, these surrogates face their own challenges: they require careful handling of the wide dynamic range of transport fluxes, must generalize across diverse plasma regimes, and need sufficient training data that adequately covers the high-dimensional input space, motivating data-efficient training strategies.

We focus on core turbulent transport surrogates using \base~\cite{meneghini2017self, Neiser22} as our baseline.

\paragraph{Bayesian Methods for Fusion.}
Bayesian methods are proven useful in fusion diagnostics and equilibrium inference~\cite{fischer2002thomson,svensson2003integrated,svensson2004integrating,svensson2008current,ford2010tokamak,kwak2020bayesian,kwak2022bayesian} and coupling between different physics components~\cite{rodriguez2024enhancing}. Beyond inference tasks, Bayesian reasoning also provides a principled framework for data acquisition when labeled samples are scarce. A closely related sub-thread targets data-efficient training of turbulent-transport surrogates through Bayesian active learning (BAL). Pool-based BAL was first applied to QuaLiKiz neural-network surrogates in~\cite{van2020fast}, and physics-informed acquisition strategies subsequently demonstrated up to $20\times$ reductions in required training data~\cite{zanisi2024efficient}. Radius-based candidate proposing has been used to detect and correct oversampled regions of the input domain during surrogate construction~\cite{kremers2023two}, while uncertainty-aware neural-network architectures paired with active learning have pushed data efficiency further~\cite{ho2025efficient}. Gaussian-process surrogates with active learning have shown complementary promise for gyrokinetic modeling of micro-tearing modes~\cite{hornsby2024gaussian}. Our work builds on this line of active-learning-assisted surrogate training; the specific BAL formulation used here is detailed in Section~\ref{sec:method:BAL}.

\section{Preliminaries}
\subsection{TGLF and Saturation Rules}
\label{sec:tglfnn:sat}
TGLF computes the total turbulent fluxes of particles \(\Gamma\), momentum \(\Pi\), and heat \(Q\) by summing the contributions from each linear mode at different wavenumbers \(\{k_{y}^1, \dots, k_{y}^{N_y}\}\):
\begin{equation}
    \Gamma = \sum_{j=1}^{N_y} \Gamma^j(k_{y}^j), \quad
    \Pi = \sum_{j=1}^{N_y} \Pi^j(k_{y}^j), \quad
    Q = \sum_{j=1}^{N_y} Q^j(k_{y}^j).
    \label{eq:tglf_summed_fluxes}
\end{equation}
Here, \(\Gamma^j, \Pi^j, Q^j\) denote the corresponding fluxes at each wavenumber \(k_{y}^j\). For conciseness, we omit the subscripts \(i\) or \(e\) in \eqref{eq:tglf_summed_fluxes} indicating particle contributions (e.g., from ions or electrons); these are reintroduced in later result sections when metrics require distinguishing between particle types. These contributions are obtained by first solving reduced-order model (ROM) equations at thermal equilibrium to compute quasilinear weights and initial intensities. The intensities are then adjusted using ``saturation rules''~\cite{waltz1999ion,kinsey2005predicting,staebler2007theory} to account for nonlinear interactions between linear modes at saturation states. Finally, the per-wavenumber fluxes \(\Gamma^j, \Pi^j, Q^j\) are computed by multiplying the quasilinear weights by the adjusted intensities.
We note that Equation~\eqref{eq:tglf_summed_fluxes} uses simplified notation: while the saturated fluxes at each $k_y^j$ are written as functions of that wavenumber alone, the saturation rules (particularly SAT2/SAT3) introduce cross-scale coupling so that the intensity at one wavenumber depends on contributions from non-local wavenumbers. Our surrogate learns these post-saturation flux contributions as training targets; the cross-scale interactions are therefore already embedded in the data (see Section~\ref{sec:spectral_regularization}).

We adopt the standard SAT2 wavenumber grid of $N_y = 24$ values~\cite{Staebler20}, constructed from local plasma quantities (ion gyroradius and flux-surface geometry) so that the $\{k_y^j\}$ differ slightly from sample to sample. Our surrogate therefore treats $k_y$ as a continuous input and is trained on a distribution of grids, which motivates the shared-weight 24-branch architecture detailed in Section~\ref{sec:spectral_regularization}.
For a complete derivation of the TGLF model, we kindly refer the reader to \cite{staebler2005gyro}, and for the specific saturation rules used in this work, see \cite{dudding2022new}.

\subsection{\base}
\label{sec:vanilla_tglfnn}
\base~\cite{meneghini2017self, Neiser22}, the state-of-the-art (SOTA) surrogate model, is an encoder-ResNet-decoder neural network. It takes 31 physical input parameters that TGLF depends on (e.g., normalized gradients, safety factor, effective charge, shaping parameters, etc; see \cite{meneghini2017self, Neiser22} for the full list) and predicts four outputs: electron particle fluxes \(\channelone\), ion momentum fluxes \(\channeltwo\), and electron and ion heat fluxes \(\channelthree\) and \(\channelfour\).

The vanilla \base~adopts an uncommon loss treatment (with respect to the ML community) to account for the huge variation in flux magnitudes. Given a batch of predictions and targets, both are first transformed via an $\asinh$ function to squash the wide range of flux values locally. Then, a custom batch normalization strategy (where the batch mean adds some random shifts) is applied. Finally, an \(R^2\)-type loss between the processed predictions and targets is calculated for gradient computation. An ensemble of 30 models is required to produce stable inferences. We note, via preliminary tests, that this approach is only stable with very small batch sizes; combined with the ensemble requirement, this contributes to slow training, especially on large datasets. We refer interested readers to the vanilla \base~implementation (as of this date) at the GA repository\footnote{\url{https://github.com/ProjectTorreyPines/TGLFNN.jl/tree/1f080a901ee9cacf58e0d63a65a9455b5ae76e6f}}.

\section{Methodology}
\begin{figure}[!h]
    \centering
    \includegraphics[width=0.6\textwidth]{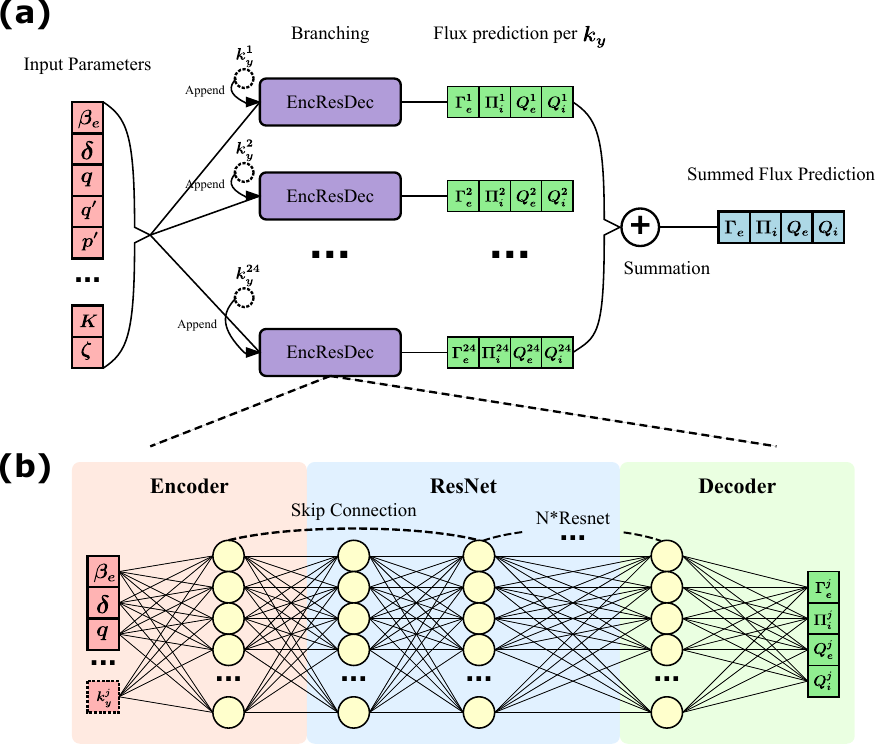}
    \caption{\textbf{Overview of \ours~architecture.} (a) \ours~takes 31 input features (the same as \base) and splits into 24 branches. Each branch produces 4 flux predictions corresponding to a specific wavenumber. These fluxes are then summed to yield the final fluxes. (b) The Encoder-ResNet-Decoder module shared by all prediction branches.}
    \label{fig:architecture_diagram}
\end{figure}

In this section, we present the methodology for implementing \ours, a surrogate model that learns the transported fluxes of TGLF. Figure~\ref{fig:architecture_diagram} (a) provides an overview of \ours: it takes the same 31 physical parameters as \base, then splits into 24 parallel branches, each appending the corresponding wavenumber to the input before predicting the transported fluxes contributed by this wavenumber. All 24 branches share learnable parameters and receive $k_y$ as a continuous input; the branch count of 24 follows standard TGLF practice for sufficient spectral resolution. Figure~\ref{fig:architecture_diagram} (b) illustrates the shared backbone for predicting fluxes at each wavenumber, utilizing an Encoder-Resnet-Decoder structure.
The shared, $k_y$-continuous mapping across branches is verified in Section~\ref{sec:res:overall} via a wavenumber-grid generalization ablation.

The remainder of this section is organized as follows. Section~\ref{sec:dataset} describes the dataset generation and filtering process. Section~\ref{sec:flux_transformation} introduces the principled feature engineering techniques employed. Section~\ref{sec:spectral_regularization} explains the implementation of wavenumber regularization in the loss term. Finally, Section~\ref{sec:method:BAL} discusses the integration of Bayesian active learning (BAL) with the training pipeline to improve data efficiency.

\subsection{Dataset}
\label{sec:dataset}
Our dataset is generated by running the numerical solver of TGLF over a dense sample of the full 31-dimensional input parameter space, denoted as \(\vx\), as described in Section~\ref{sec:tglfnn:sat}. We generate the datasets with two levels of perturbations. The first level is the \textbf{\MaPert} set, which consists of approximately 330{,}000 samples that perturb key physical parameters in a continuous, dense fashion. The second level is the \textbf{\MaMiPert} set, which builds on the \MaPert set by introducing 9 additional discrete variations, resulting in around 3 million total samples.

\paragraph{Filtering Outliers.}
Some input combinations and output solutions are rarely encountered in real experiments. Standalone TGLF simulations sample the input hypercube uniformly, producing parameter combinations that violate physical feasibility constraints and would never arise in actual plasma discharges; examples include correlations between pressure gradients and between temperature and density gradients that are imposed by collisional exchange and actuator limitations. Additionally, the numerical solver, particularly the approximate eigenvalue solver using Hermite basis functions, can converge to sub-optimal solutions due to truncation or numerical errors. As a result, the generated dataset inherently contains noise or outliers that can destabilize training. To address this, we iteratively apply a Median Absolute Deviation (MAD) filtering approach. Specifically, we compute:
\begin{equation}
    \vm = \mathrm{med}\left(\left\{\left|\vx - \tilde{\vx}\right|, \quad \text{for   } \vx \in \sX\right\}\right),
\end{equation}
where \(\tilde{\vx} = \mathrm{med}(\sX)\) is the median of the dataset \(\sX\), and \(\vm\) is the median absolute deviation (MAD). We then define a distance for each data point \(\vx\) as:
\begin{equation}
    \vd = \kappa \frac{|\vx - \tilde{\vx}|}{\vm},
\end{equation}
where all operations (absolute values and divisions) are applied element-wise across the vector components, and \(\kappa = 1.4826\) is a scaling factor that approximates the standard deviation under normality assumptions. In each iteration, any sample with \(\vd > \epsilon\) in any channel is removed, where \(\epsilon = 200\) is a threshold hyperparameter. The iteration terminates when fewer than \(1\%\) of the data points are removed in a single iteration.

The original \base~is trained on the \MaMiPert set with outliers filtered out. Unless otherwise noted, we preserve this setting in our experiments. For the ablation study on training under sparse, noisy datasets, see Section~\ref{sec:res:sparse}.

\subsection{Feature Tuning}
\label{sec:flux_transformation}
Compared to the batch-wise non-standardized treatment in \cite{meneghini2017self, Neiser22} and the mixture-of-experts (MoE) scheme in \cite{van2020fast,fransson2023fast}, we adopt a well-established feature engineering approach based on the inverse hyperbolic sine transformation~\cite{johnson1949systems, burbidge1988alternative} to handle the large variation in flux magnitudes. Let \(\vy = \{\channelone, \channeltwo, \channelthree, \channelfour\}\) denote the concatenated fluxes and \(\vy^j = \{\channelone^j(k_y^j), \channeltwo^j(k_y^j), \channelthree^j(k_y^j), \channelfour^j(k_y^j)\}\) denote the contributions at the \(j\)-th wavenumber for \(j = 1, \dots, N_y\). We apply $\asinh$ transformations to both the final flux predictions \(\vy' = (\mathcal{S} \circ \asinh)(\vy)\) and the per-wavenumber contributions \(\vy'^j = (\mathcal{S} \circ \asinh)(\vy^j)\), where \(\mathcal{S}\) represents standardization across the training dataset. The choice of $\asinh$ over $\log$ is motivated by the need to accommodate possible negative predictions in TGLF.

\subsection{Wavenumber Regularization}
\label{sec:spectral_regularization}
Building on the per-wavenumber predictions $f_\theta(\vx, k_y^j)$ introduced in Figure~\ref{fig:architecture_diagram}, we define the supervised loss on the turbulent transport flux as:
\begin{equation}
    \mathcal{L}_{f} = \|\vy' - \sum_{j=1}^{N_y} f_{\theta}\left(\vx, k_y^j\right) \|^2.
\end{equation}
In addition, we impose a regularization term \(\mathcal{L}_{s}\) on the per-wavenumber fluxes:
\begin{equation}
    \mathcal{L}_{s} = \sum_{j=1}^{N_y} \| \vy'^j - f_{\theta}\left(\vx, k_y^j\right) \|^2.
\end{equation}
This dual supervision acts as a physics-informed regularization that constrains the model to learn physically meaningful per-wavenumber decompositions. While a perfect model would theoretically minimize both losses simultaneously, in practice this regularization provides additional constraints that improve generalization, especially with limited training data, as demonstrated in our ablation studies (Section~\ref{sec:res:overall}).
Combined, the overall objective function reads:
\begin{equation}
    \mathcal{L} = \mathcal{L}_{f} + \mathcal{L}_{s}.
    \label{eq:combined_loss}
\end{equation}

\subsection{Bayesian Active Learning}
\label{sec:method:BAL}
To study how the model performs under sparse datasets and active learning, particularly in preparation for learning higher-fidelity models, we employ Bayesian Active Learning (BAL) in a pool-based setting. The pool-based approach is chosen because the full dataset is available in our case, though the method can be minimally modified for online settings. BAL improves data efficiency by strategically selecting the most informative data points for training the surrogate model, reducing the amount of data to be simulated.

The BAL process proceeds as follows. We start with a small initial labeled dataset \(\mathcal{D}_0\). At each iteration \(t\), we apply an acquisition function \(g(\vx)\) to rank all candidates in the unlabeled input space \(\mathcal{U}\). The top \(k\) input candidates are selected based on their acquisition scores, labeled by running the TGLF numerical simulation \(\vy=L(\vx)\), and added to \(\mathcal{D}_t\) to form an augmented dataset \(\mathcal{D}_{t+1}\). The model is then retrained on \(\mathcal{D}_{t+1}\), and its performance is reevaluated. This iterative process continues until a stopping criterion is met. The dataset update at iteration \(t\) reads:
\begin{equation}
    \mathcal{D}_{t+1} = \mathcal{D}_t \cup \{(\vx, L(\vx)) \mid \vx \in \text{Top-}k(g(\vx') \text{ for all } \vx' \in \mathcal{U})\}.
    \label{eq:bal}
\end{equation}

Our primary study target is the \emph{Expected Information Gain} (EIG) acquisition function, while random selection (i.e., sampling using the full dataset's distribution) serves as a natural baseline for comparison. In a pool-based BAL, the chosen candidates are finally mapped to their nearest neighbor in the dataset.

\paragraph{Expected Information Gain.}
The EIG acquisition function estimates how much information is gained, i.e., the reduction in entropy, by acquiring new data entries. For each candidate \(\vx\), we compute the \emph{prior entropy} \(H[p(\vy' \mid \vx)] = -\sum_i p(\vy'_i \mid \vx) \log p(\vy'_i \mid \vx)\)~\cite{shannon1948mathematical} based on the variance of the model's predictions for \(\vx\). The \emph{posterior entropy} is obtained using an updated model, which is retrained on an intermediate-augmented dataset \(\mathcal{D}_t^{*} = \mathcal{D}_t \cup \{(\vx, \vy')\}\), i.e., taking all proposed candidates while treating their \(\vy'\) as ground truth. To estimate model uncertainty given our deterministic neural network structure, we incorporate a dropout layer during inference and ensemble over $K=16$ stochastic forward passes, following the Monte Carlo Dropout framework~\cite{gal2016dropout} which interprets dropout at test time as approximate Bayesian inference. The EIG formally reads:
\begin{equation}
    g_{\text{EIG}}(\vx) = H[p(\vy' \mid \vx)] - \mathbb{E}_{p(\theta^{*} \mid \mathcal{D}_t^{*})}[H[p(\vy'^{*} \mid \vx, \theta^{*})]],
    \label{eq:bal:eig}
\end{equation}
where \(\theta^{*}\) and \(\vy'^{*}\) are the parameters of the neural network and the prediction after training on the intermediate-augmented dataset \(\mathcal{D}_t^{*}\).

\paragraph{Radius-Based Candidate Proposing.}
To propose candidates for EIG, we sample randomly according to the input variables' distribution at different minor radius locations, which helps conserve the variation of local distribution in the transport. Specifically, we partition the radial domain into brackets and compute the per-bracket mean and standard deviation for each of the 31 input parameters from the full dataset metadata (see Appendix~\ref{sec:appendix:RP}). Candidates are then drawn from Gaussian distributions parameterized by these per-radius statistics, preserving the local correlations between input parameters that characterize different radial positions. This physics-informed proposing ensures that acquisition candidates reflect realistic parameter combinations at each radius, rather than sampling uniformly from the global hypercube. An ablation study on the use of this prior distribution information is provided in Section~\ref{sec:res:bal}.

\section{Experimental Results}
This section presents the experimental results of our study. The remainder of this section is organized as follows:
Section~\ref{sec:res:overall} compares the offline performance of the existing state-of-the-art (SOTA) \base~with our new \ours. The contributions of individual components proposed are further analyzed.
Section~\ref{sec:res:sparse} examines the advantages of \ours~under sparse and noisy datasets, specifically using only the \MaPert~set without outlier filtering.
In Section~\ref{sec:res:bal}, we report the comparison results for Bayesian Active Learning (BAL) under three different controlled settings.
Finally, Section~\ref{sec:res:fluxmatching} uses a case study to verify the effectiveness of \ours~in downstream tasks, specifically the flux-matching problem in tokamak design workflows.

For brevity, supporting results are collected in the appendices: the complete ablation table (Appendix~\ref{sec:appendix_ablation}), the hyperparameter search (Appendix~\ref{sec:appendix:hyperparam}), per-branch parity analysis (Appendix~\ref{sec:appendix_per_branch_parity}), BAL dataset-distribution analysis (Appendix~\ref{sec:appendix_bal_dist}), TGLF fluxes evaluated at NN-converged profiles (Appendix~\ref{sec:appendix_tglf_at_nn}), stored thermal energy sanity check (Appendix~\ref{sec:appendix:wth_lumped}), aggregated validation across the large DIII-D dataset (Appendix~\ref{sec:appendix:aggregated_validation}), and BAL computational timing (Appendix~\ref{sec:appendix:bal_timing}).

For evaluation metrics, we mainly report the \textbf{Root Mean Squared Logarithmic Error (RMSLE)} to account for the large variation in flux magnitudes. The RMSLE is defined as:
\begin{equation}
    \text{RMSLE}(\vy', \vy) = \sqrt{\frac{1}{N} \sum_{i=1}^{N} \left\|\ln\left(1 + \left|\vy'_i\right|\right) - \ln\left(1 + \left|\vy_i\right|\right)\right\|^2},
    \label{eq:rmsle}
\end{equation}
where \(N\) is the number of test samples, \(\vy'_i\) represents the predicted fluxes, and \(\vy_i\) represents the TGLF target fluxes.

For comparison, we present results and figures for all transport channels. This approach prevents potentially masking poor performance in any individual channel, while demonstrating that accuracy improvements are consistently observed across all physical variables. We provide a detailed channel-specific analysis in Section~\ref{sec:res:fluxmatching} for the convergence behavior and performance differences between individual transport channels during flux matching.

\subsection{Overall Accuracy Improvements}
\label{sec:res:overall}
\begin{figure}[!h]
    \centering
    \begin{subfigure}[c]{0.23\textwidth}
        \centering
        \includegraphics[width=\textwidth]{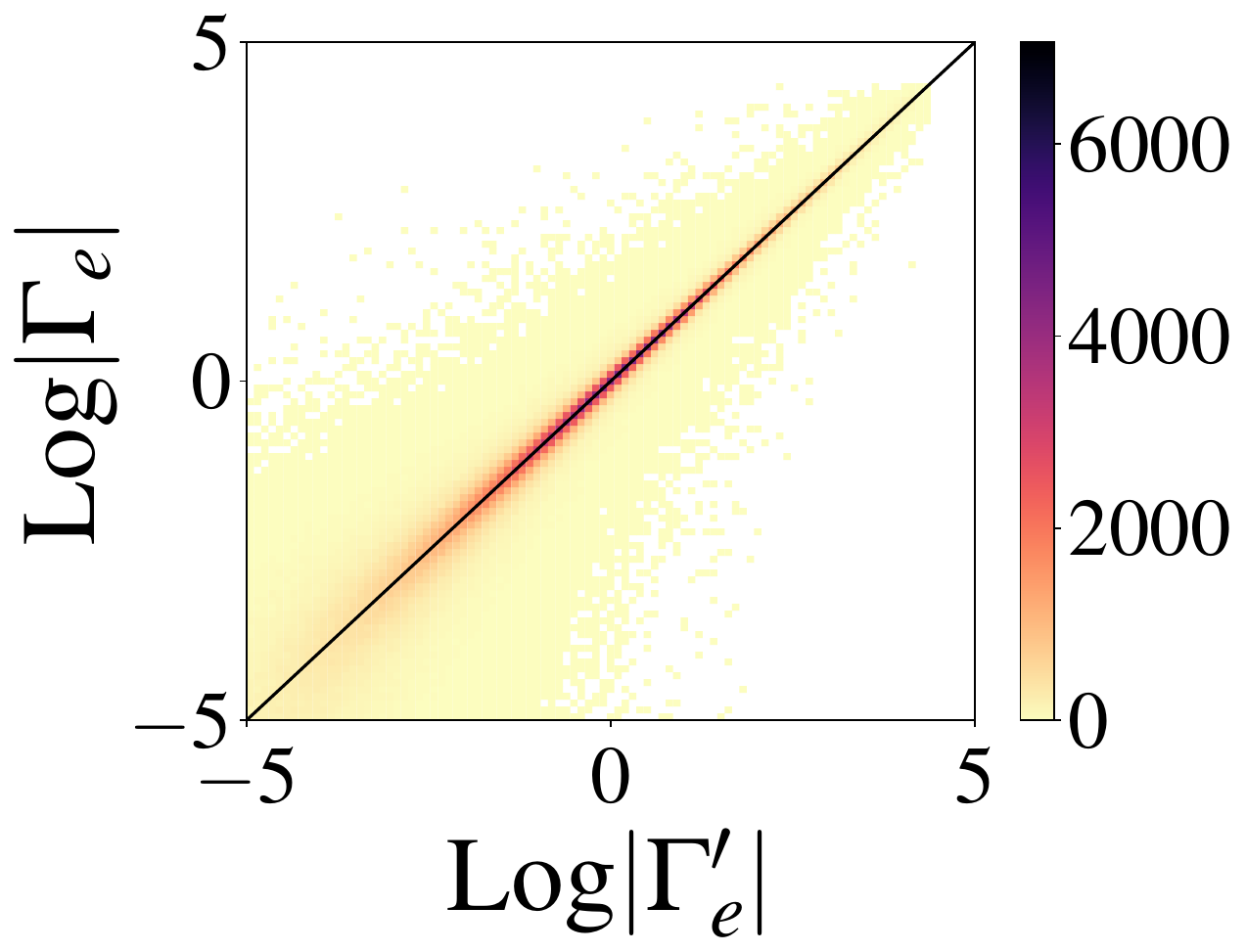}
    \end{subfigure}
    \hfill
    \begin{subfigure}[c]{0.23\textwidth}
        \centering
        \includegraphics[width=\textwidth]{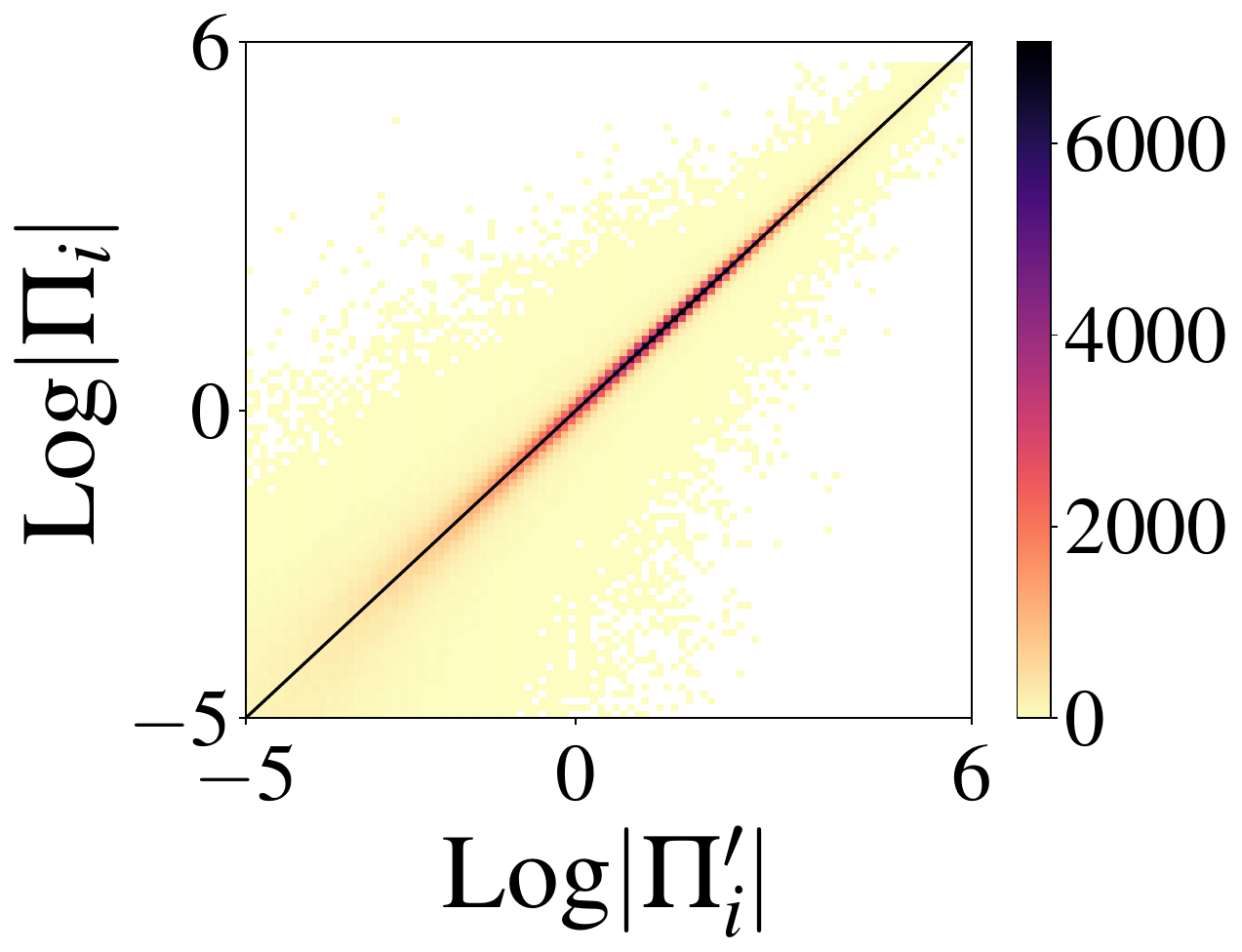}
    \end{subfigure}
    \hfill
    \begin{subfigure}[c]{0.23\textwidth}
        \centering
        \includegraphics[width=\textwidth]{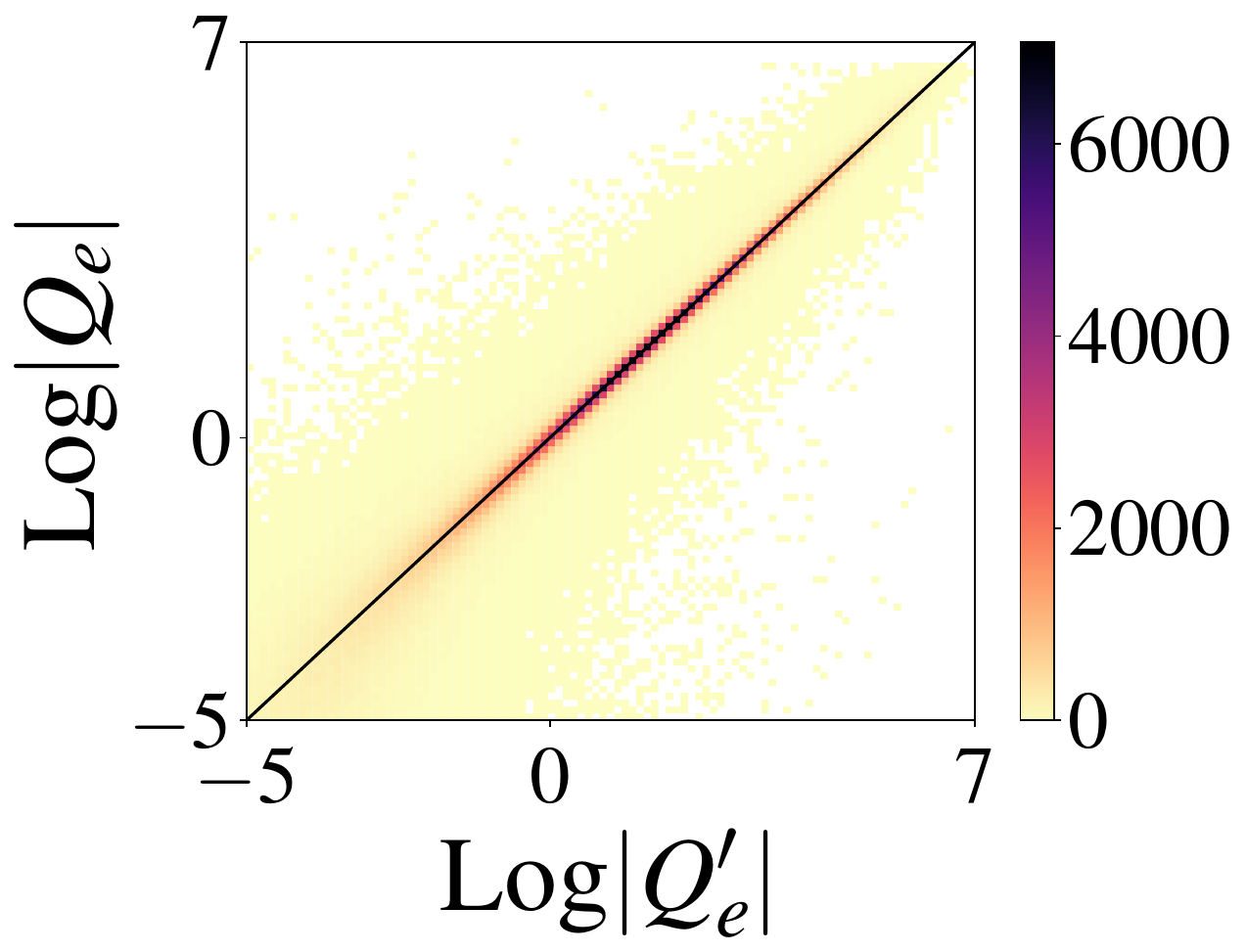}
    \end{subfigure}
    \hfill
    \begin{subfigure}[c]{0.23\textwidth}
        \centering
        \includegraphics[width=\textwidth]{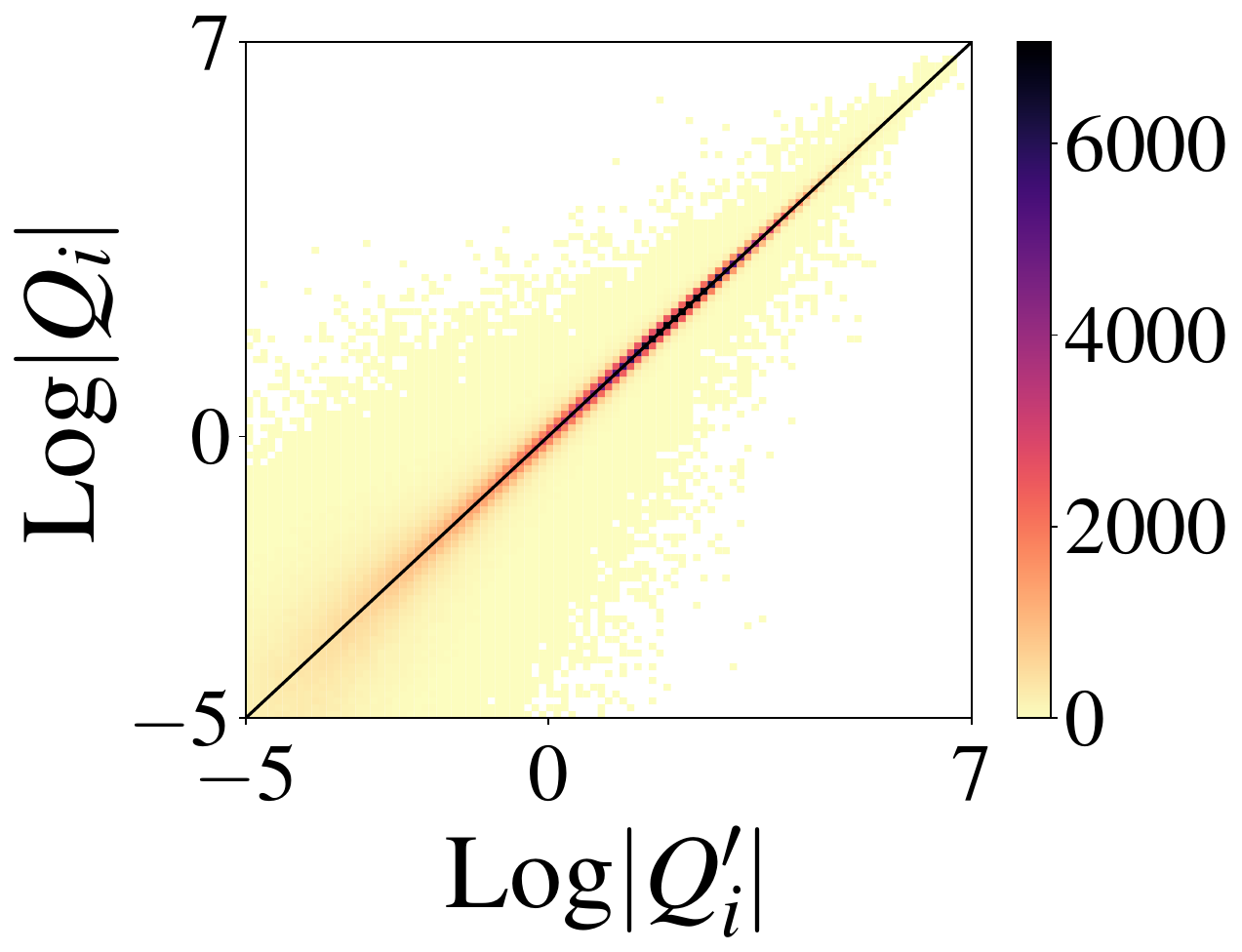}
    \end{subfigure}

    \caption{\textbf{Prediction accuracy of \ours}. Heatmaps compare predicted fluxes (Pred) $\vy'$ against target fluxes $\vy$ for electron particle flux ($\channelone$), ion momentum flux ($\channeltwo$), electron heat flux ($\channelthree$), and ion heat flux ($\channelfour$). The diagonal line represents perfect agreement. Color intensity indicates sample density, with darker regions representing higher concentrations of data points.}
    \label{fig:overall:recon}
\end{figure}

We first demonstrate the qualitative accuracy of \ours~by presenting heatmaps of predicted versus target fluxes in Figure~\ref{fig:overall:recon}. The results indicate that \ours~accurately predicts fluxes across a wide range of plasma conditions, with most data points closely aligned along the diagonal, which indicates strong agreement between the predictions and the TGLF targets.

\begin{figure}[!h]
    \centering
    \includegraphics[width=\textwidth]{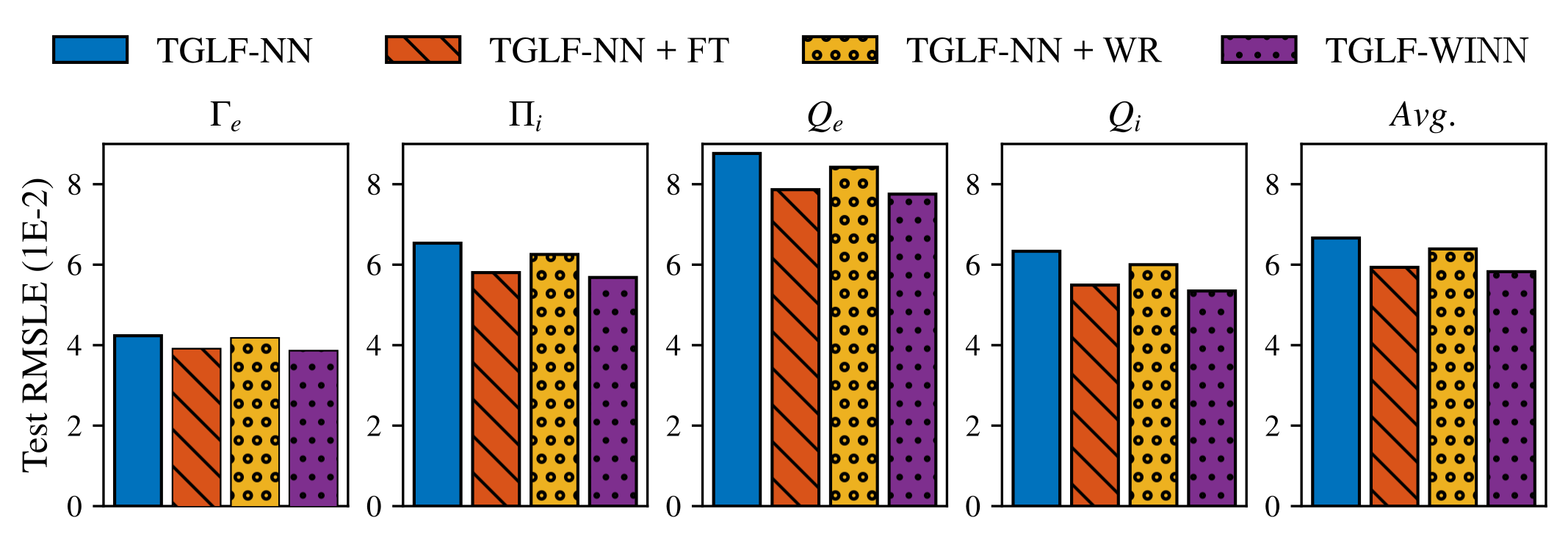}
    \caption{\revision{\textbf{Offline error comparison across different models.}} Root Mean Squared Logarithmic Error (RMSLE) ($\times 10^{-2}$) is used as the metric. The graph compares errors for all four flux channels and their average over the full dataset, consistent with~\cite{meneghini2017self, Neiser22}. Lower values indicate superior performance. Abbreviations: \firstcontri = Feature Tuning, \secondcontri = Wavenumber Regularization.}
    \label{fig:overall}
\end{figure}

We then present a comprehensive comparison of offline performance between \ours~and \base, on the consistent dataset as in~\cite{meneghini2017self, Neiser22} (ie, the set \MaMiPert~with outliers being filtered). To isolate the contributions of our innovations, we also evaluated the performance of adding either feature tuning or wavenumber regularization to the original \base~as ablation studies, as shown in Figure~\ref{fig:overall}.

Our method reduces the RMSLE from \(6.66 \times 10^{-2}\) to \(5.83 \times 10^{-2}\), achieving a relative improvement of 12.5\%. The introduction of feature tuning and wavenumber regularization yields relative improvements of 11.0\% and 4.1\%, respectively.

As shown in Figure~\ref{fig:overall}, both feature tuning (\firstcontri) and wavenumber regularization (\secondcontri) contribute complementary benefits: \firstcontri~improves accuracy by compressing the target prediction range via the $\asinh$ transformation and standardization, while \secondcontri~provides physics-guided decomposition constraints that regularize per-wavenumber predictions.

Complete per-channel RMSLE values are reported in Appendix~\ref{sec:appendix_ablation}, Table~\ref{tab:ablation_results}.

Per-branch parity plots across all 24 branches (Appendix~\ref{sec:appendix_per_branch_parity}) confirm that $\mathcal{L}_s$ (Eq.~\ref{eq:combined_loss}) constrains individual branches rather than only the summed flux.

We further verify that the shared-weight branches represent a $k_y$-continuous mapping rather than 24 independently-tuned predictors. Each test sample's original per-branch wavenumbers $\{k_y^j\}_{j=1}^{24}$ were replaced by the dataset-mean grid $\{\bar{k}_y^j\}_{j=1}^{24}$ (the per-branch average of $k_y^j$ across the training set), so that every test sample is queried on a single shared grid never seen exactly during training. The offline RMSLE increases by $3.02\times 10^{-2}$ (from $5.83\times 10^{-2}$ to $8.85\times 10^{-2}$), confirming that the network produces coherent predictions when queried on a novel $k_y$ grid without any retraining.

\subsection{Robustness against Sparsity and Noisy Data}
\label{sec:res:sparse}
\begin{figure}[!h]
    \centering
    \includegraphics[width=1.0\textwidth]{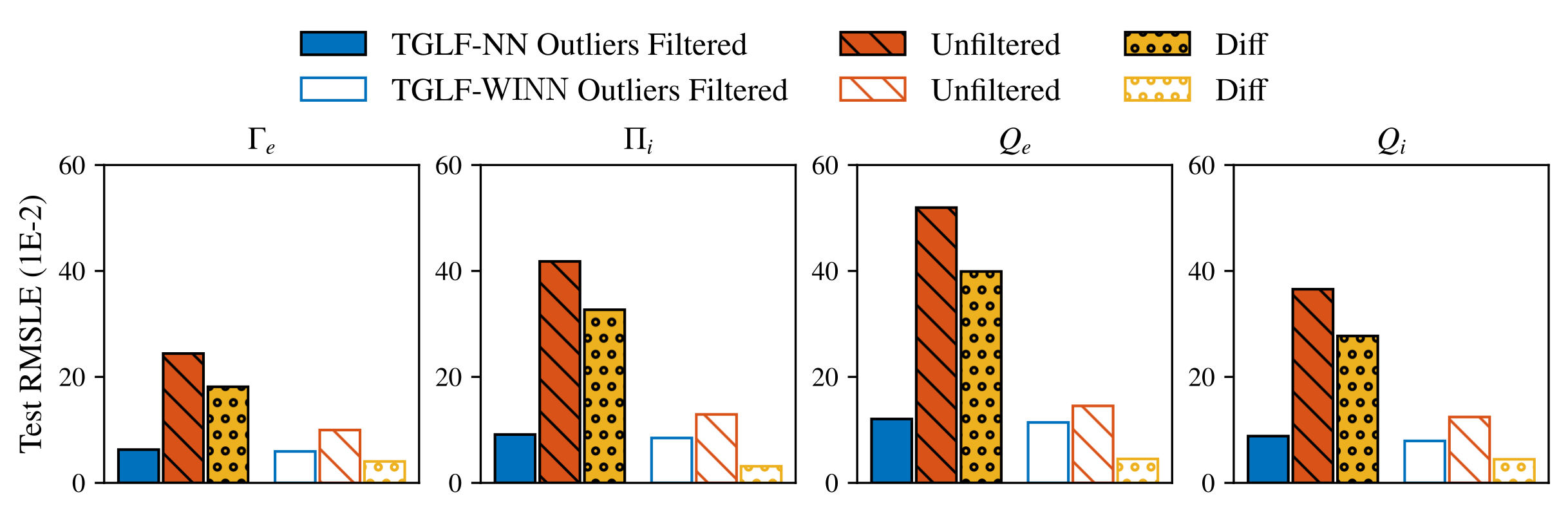}
    \caption{\revision{\textbf{Comparison of \base~and \ours~performance under a sparse dataset}.} Models are compared with and without outlier filtering. Metrics are reported in Root Mean Squared Logarithmic Error (RMSLE) ($\times 10^{-2}$) for each channel. Lower values indicate better performance.}
    \label{fig:compare_filtered}
\end{figure}

Generating a full dataset is prohibitively expensive, especially if one wishes to adapt our method to higher-fidelity gyrokinetic simulations~\cite{Neiser23,Lestz25,Patel24}. Additionally, without a complete dataset, the filtering procedure described in Section~\ref{sec:dataset} becomes impractical. Thus, it is essential to evaluate training methods under sparse and noisy conditions.
To evaluate robustness under data sparsity, we use the \MaPert~dataset alone (see Section~\ref{sec:dataset}), which contains approximately 330{,}000 samples, only 1/9 the size of the full \MaMiPert~dataset ($\sim$3 million samples). We compare the offline performance of \base~and \ours~with and without outlier filtering (see Section~\ref{sec:dataset} for details on MAD filtering). The results are shown in Figure~\ref{fig:compare_filtered}.

\ours~consistently outperforms \base~across all settings. While both methods achieve similar RMSLE values after filtering (\(8.66 \times 10^{-2}\)  vs. \(9.3 \times 10^{-2}\)), \ours~shows significantly lower RMSLE in the unfiltered setting (\(12.57 \times 10^{-2}\) vs. \(39.93 \times 10^{-2}\)), i.e., a smaller performance degradation (RMSLE increase of \(3.91 \times 10^{-2}\) for \ours~vs.~\(30.63 \times 10^{-2}\) for \base), highlighting its robustness to noise and sparsity.

\begin{figure}[!h]
    \centering
    \begin{subfigure}[c]{0.48\textwidth}
        \centering
        \includegraphics[width=\textwidth]{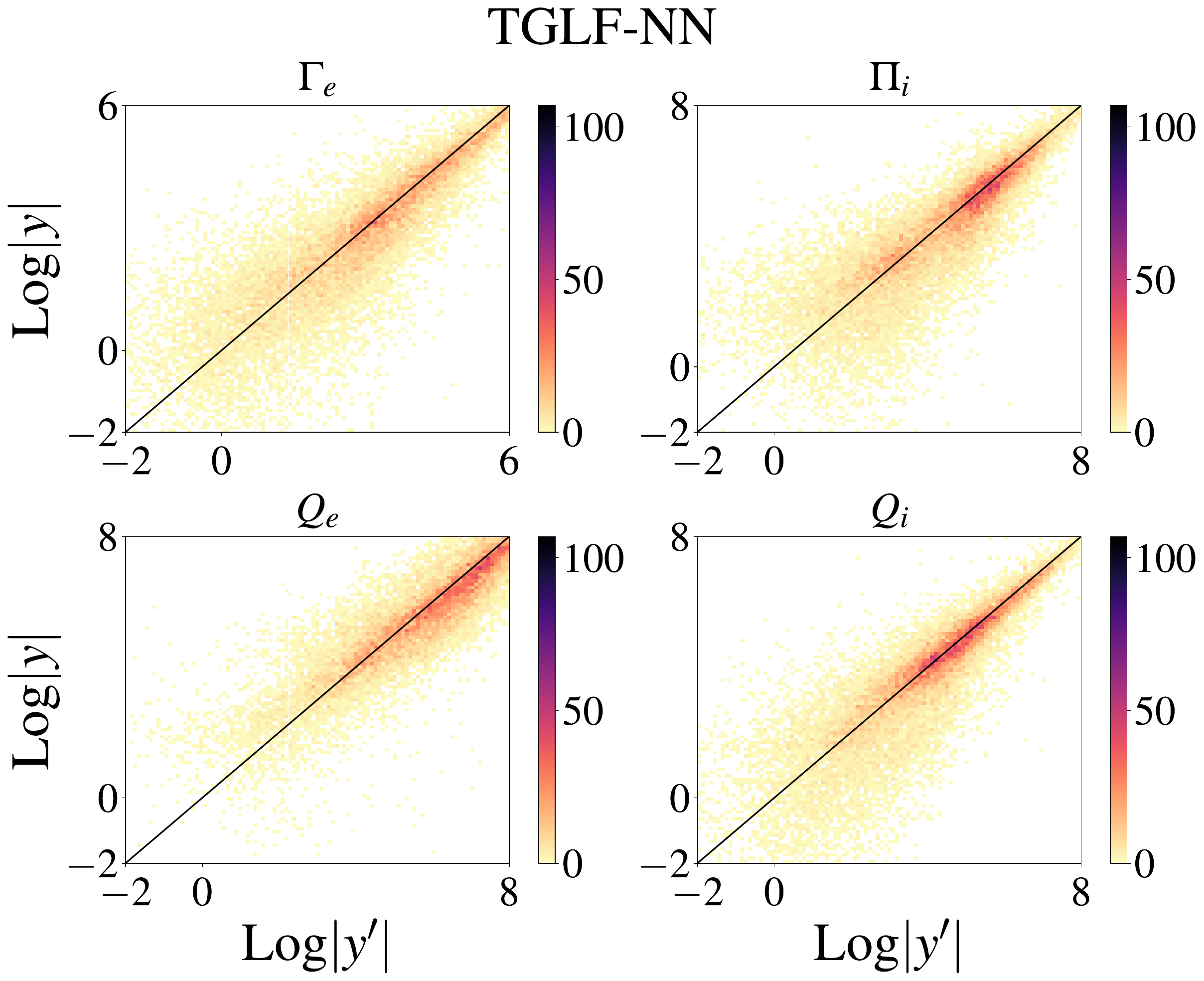}
        \label{fig:subfig1}
    \end{subfigure}
    \hfill
    \begin{subfigure}[c]{0.48\textwidth}
        \centering
        \includegraphics[width=\textwidth]{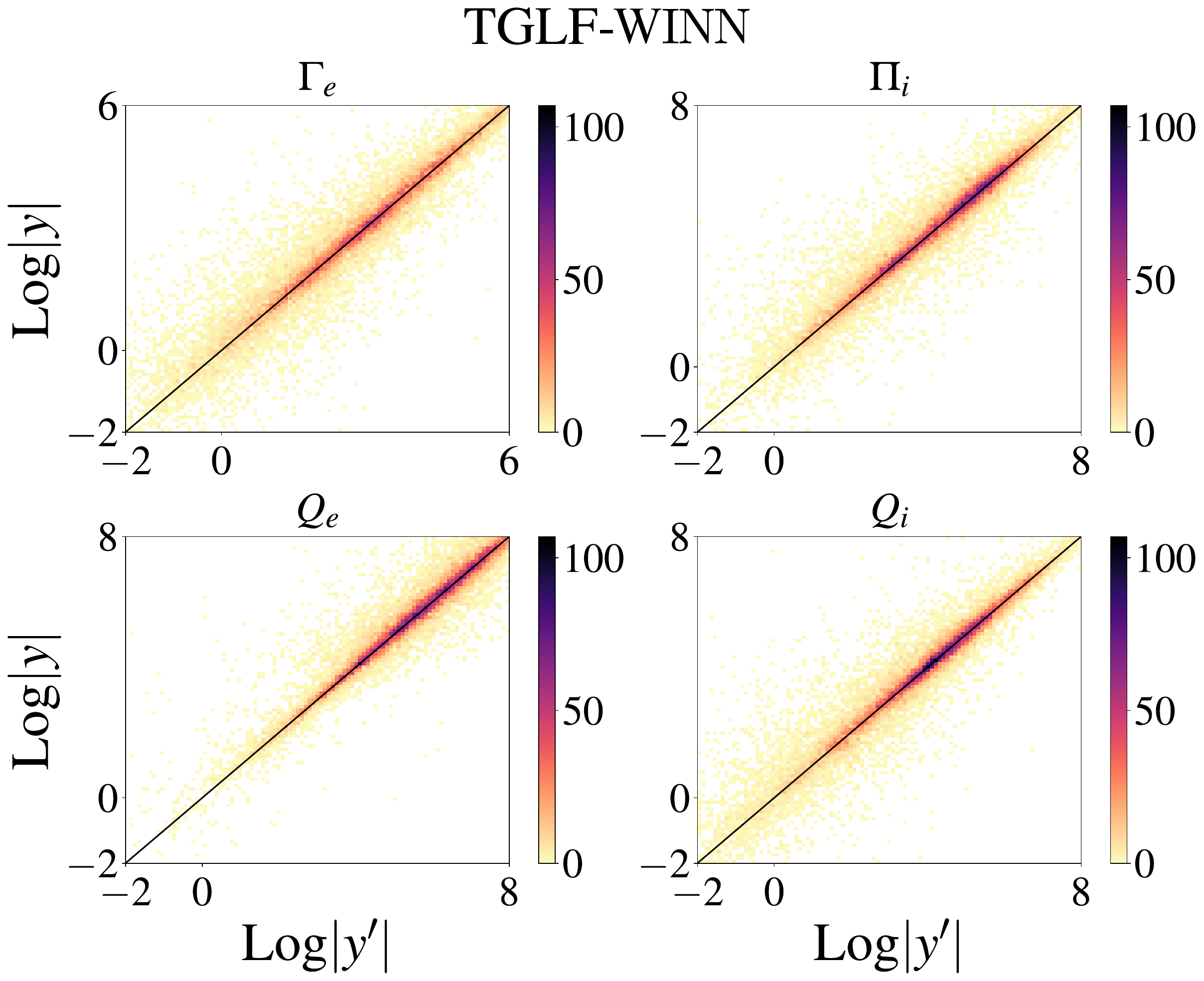}
        \label{fig:subfig4}
    \end{subfigure}

    \caption{\revision{\textbf{Heatmaps comparing prediction of \textit{removed outliers}.}} These plots compare predicted fluxes (Pred) $\vy'$ versus TGLF target fluxes $\vy$. We compare all four channels (\(\channelone\), \(\channeltwo\), \(\channelthree\), \(\channelfour\)) for \base~and \ours~over outliers in the \MaPert~dataset. The diagonal line represents perfect agreement. The colorbar represents density of samples, where darker regions have more samples.}
    \label{fig:sparse:outlier}
\end{figure}

To further analyze performance on outliers, we evaluate models that were trained on the unfiltered \MaPert~dataset, and then plot heatmaps (prediction vs. TGLF target) for the subset of the test set flagged as outliers by the MAD filter in Figure~\ref{fig:sparse:outlier}. Most outliers exhibit high flux magnitudes, consistent with the MAD filtering mechanism. \ours~demonstrates closer alignment to the diagonal, particularly in the high-magnitude range, showcasing the effectiveness of the innovations: range-squashing and regularization introduced.

\subsection{Bayesian Active Learning}
\label{sec:res:bal}
We conducted Bayesian Active Learning (BAL) experiments with three ablations: (1) model, (2) acquisition function, and (3) incorporation of physics priors or not. We aim to identify the optimal BAL configuration for \ours~and provide insights for future research on active learning for higher fidelity models. Unless otherwise specified, \textbf{our default configuration for the acquisition function is the Expected Information Gain (EIG) with per-radius distributions when proposing candidates.}

\begin{figure}[!h]
    \centering
    \begin{subfigure}{0.32\textwidth}
        \centering
        \includegraphics[width=\textwidth]{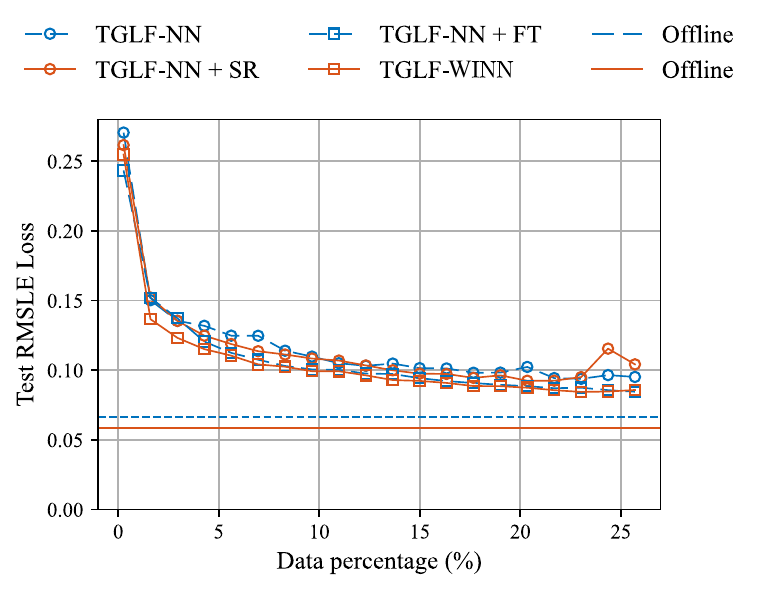}
        \caption{Model comparison}
        \label{fig:bal_compare_loss_function}
    \end{subfigure}
    \hfill
    \begin{subfigure}{0.32\textwidth}
        \centering
        \includegraphics[width=\textwidth]{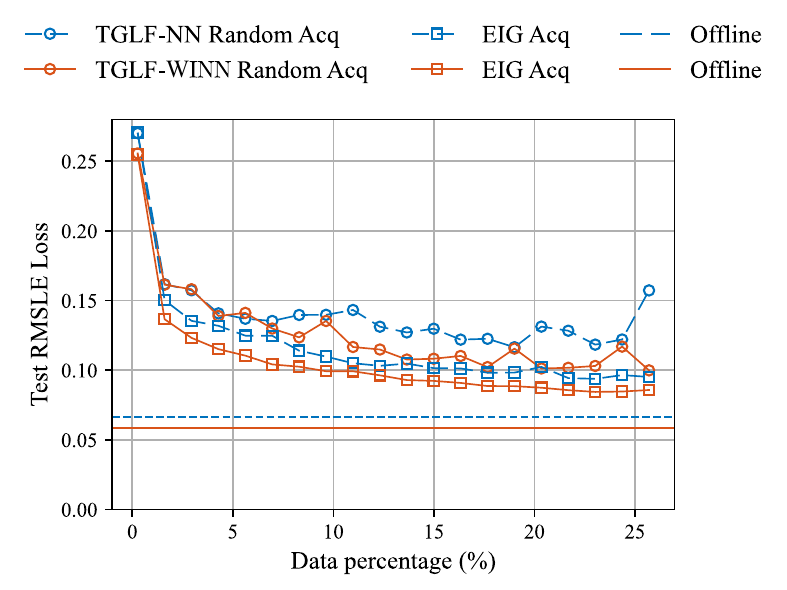}
        \caption{Acquisition function}
        \label{fig:bal_compare_acq_function}
    \end{subfigure}
    \hfill
    \begin{subfigure}{0.32\textwidth}
        \centering
        \includegraphics[width=\textwidth]{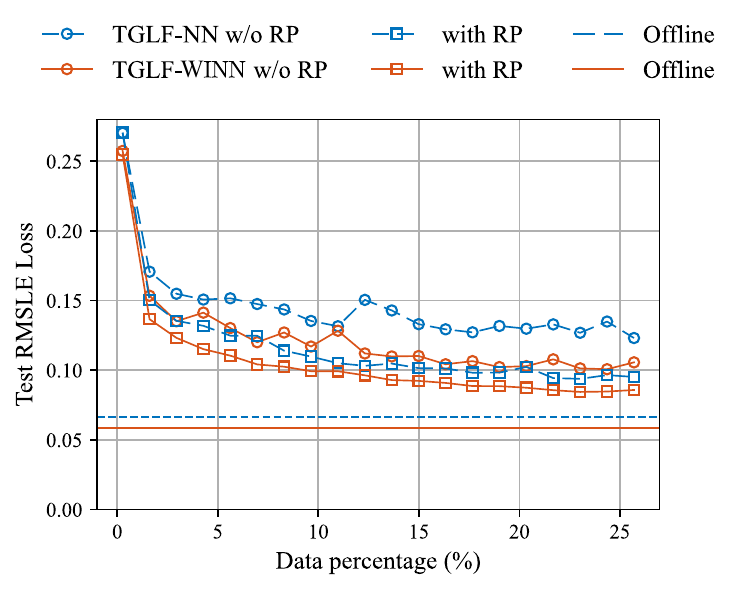}
        \caption{Radius-based proposing}
        \label{fig:bal_compare_rho}
    \end{subfigure}
    \caption{\revision{\textbf{Bayesian Active Learning ablation studies:}} (a) Error convergence comparison of \base, \base~+ \firstcontri, \base~+ \secondcontri, and \ours~showing \ours~achieves fastest error reduction. (b) EIG consistently outperforms random sampling for both models. (c) Incorporating physics priors through radius-based proposing significantly improves performance for both models.}
    \label{fig:bal_ablations}
\end{figure}

\paragraph{Model.}
Figure~\ref{fig:bal_compare_loss_function} illustrates the BAL performance of four models: \base, \base~+ \firstcontri, \base~+ \secondcontri, and \ours. The relative performance ranking among models aligns with the offline results presented in Section~\ref{sec:res:overall} and Figure~\ref{fig:overall}. \ours~exhibits the most rapid convergence, reducing errors by $0.0118$ compared to \base~at 25\% data utilization, representing a $12.40\%$ error reduction. In particular, with only 25\% of the data, \ours~achieves an error level that is merely $0.0165$ higher than \base's offline accuracy and $0.0248$ higher than our best offline model.

\paragraph{Acquisition Function.}
Figure~\ref{fig:bal_compare_acq_function} compares the performance of \base~and \ours~using two acquisition functions: Expected Information Gain (EIG) and random sampling. EIG consistently outperforms random sampling by targeting inputs that minimize entropy, thereby reducing uncertainty in critical parameter regions. At 25\% data utilization, EIG reduces errors by $0.0196$ for \ours~and $0.0249$ for \base.

\paragraph{Radius-Based Candidates Proposing.}
Physical input parameters exhibit varying distributions at different major radii. We ablate whether incorporating this knowledge when proposing candidates for EIG improves performance. Figure~\ref{fig:bal_compare_rho} demonstrates that incorporating this prior into candidate proposing reduces errors by $0.0306$ for \ours~and $0.0367$ for \base~at 25\% data utilization. A complementary marginal-distribution diagnostic (Appendix~\ref{sec:appendix_bal_dist}) confirms the BAL and random pools are nearly identical in input marginals; BAL's gain therefore comes from sample-level ordering rather than input-space coverage.

\subsection{Flux-matched profile predictions}
\label{sec:res:fluxmatching}

\begin{figure}[!h]
    \centering
    \begin{subfigure}{0.8\textwidth}
        \centering
        \includegraphics[width=\textwidth]{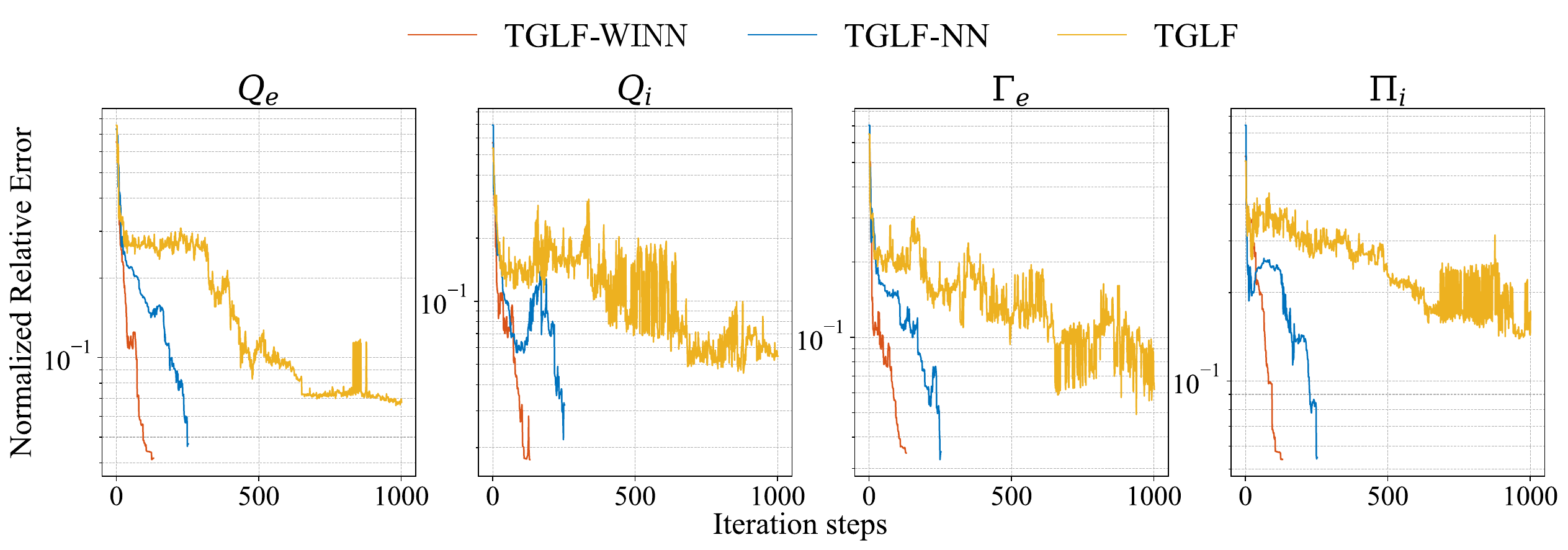}
        \caption{L-mode convergence}
        \label{fig:convergence_lmode}
    \end{subfigure}

    \begin{subfigure}{0.8\textwidth}
        \centering
        \includegraphics[width=\textwidth]{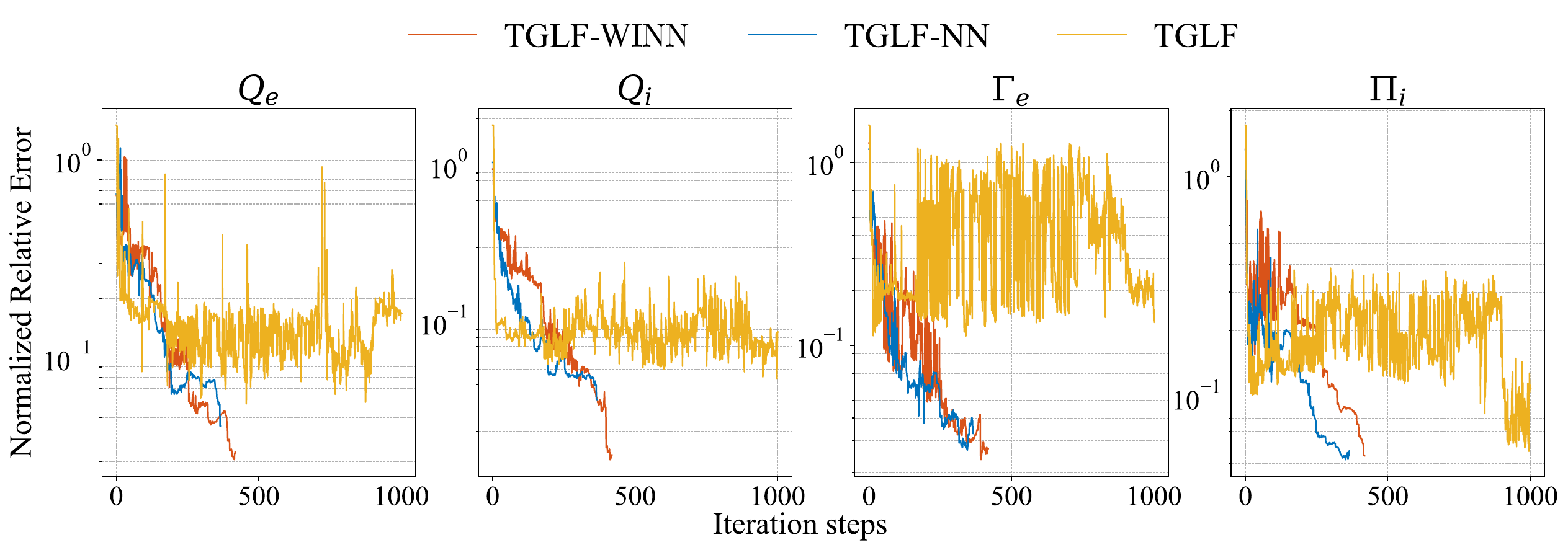}
        \caption{H-mode convergence}
        \label{fig:convergence_hmode}
    \end{subfigure}
    \caption{\revision{\textbf{Convergence of flux-matching in L-mode and H-mode scenarios}.} Comparison of the relative error between the predicted flux and the prescribed source across iteration steps for TGLF, \base, and \ours. (a) L-mode: \ours~converges after 129 iterations, \base~after 251 iterations, while TGLF fails to converge after 1000 iterations. (b) H-mode: \ours~converges after 412 iterations, \base~after 364 iterations, while TGLF fails to converge after 1000 iterations.}
    \label{fig:convergence_combined}
\end{figure}

\begin{figure}[!h]
    \centering
    \begin{subfigure}{0.8\textwidth}
        \centering
        \includegraphics[width=\textwidth]{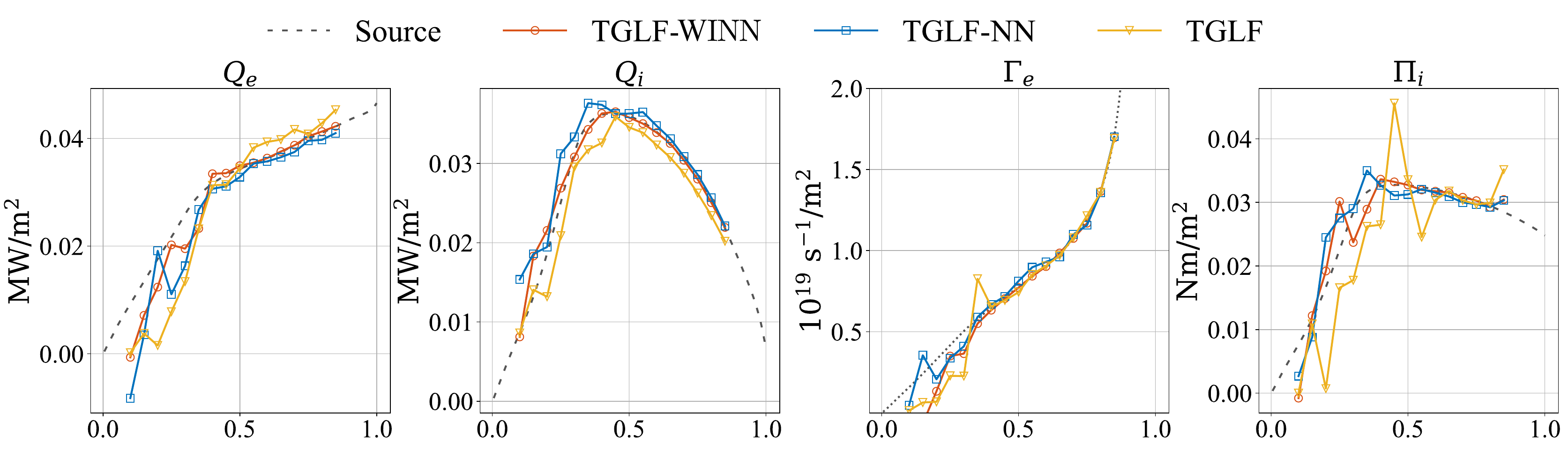}
        \caption{L-mode flux profiles}
        \label{fig:flux_profiles_lmode}
    \end{subfigure}

    \begin{subfigure}{0.8\textwidth}
        \centering
        \includegraphics[width=\textwidth]{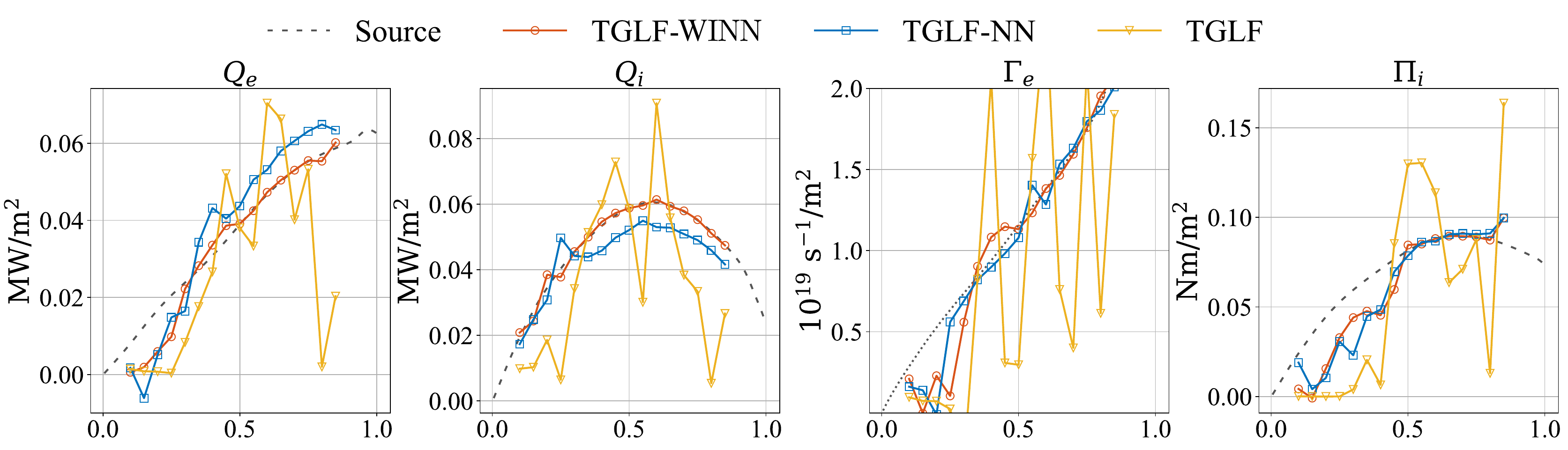}
        \caption{H-mode flux profiles}
        \label{fig:flux_profiles_hmode}
    \end{subfigure}
    \caption{\revision{\textbf{Reconstructed flux profiles in L-mode and H-mode scenarios}.} Comparison of \ours, \base, and TGLF for predicted transport fluxes ($Q_e$, $Q_i$, $\Gamma_e$, $\Pi$) over 16 $\rho$ locations. (a) L-mode reconstructed fluxes. (b) H-mode reconstructed fluxes.
    }
    \label{fig:flux_profiles_combined}
\end{figure}


\begin{figure}[!h]
    \centering
    \begin{subfigure}{0.5\textwidth}
        \centering
        \includegraphics[width=\textwidth]{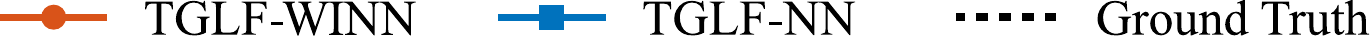}
    \end{subfigure}

    \begin{subfigure}{0.8\textwidth}
        \centering
        \includegraphics[width=\textwidth]{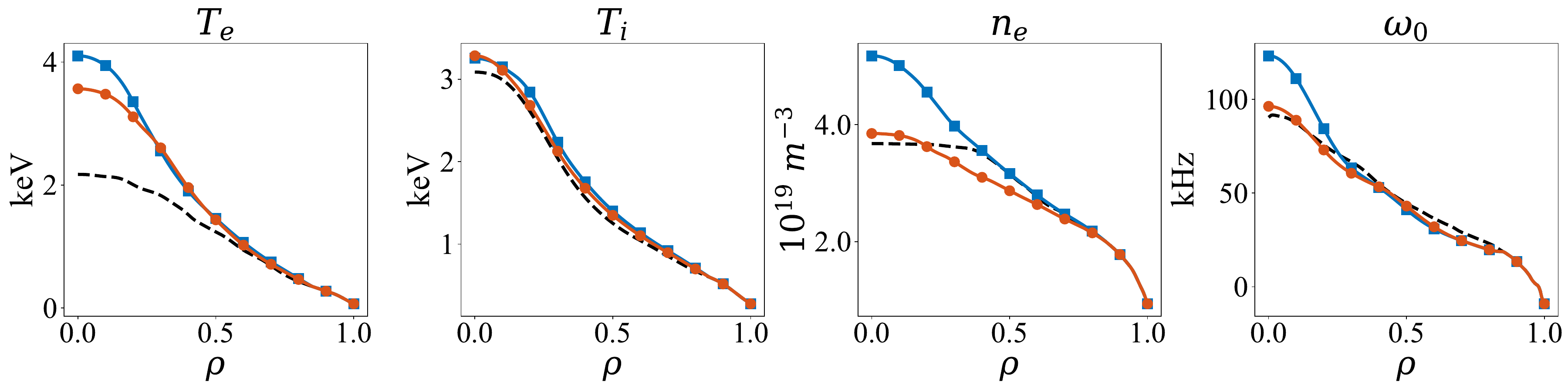}
        \caption{L-mode core profiles}
        \label{fig:profiles_lmode}
    \end{subfigure}

    \begin{subfigure}{0.8\textwidth}
        \centering
        \includegraphics[width=\textwidth]{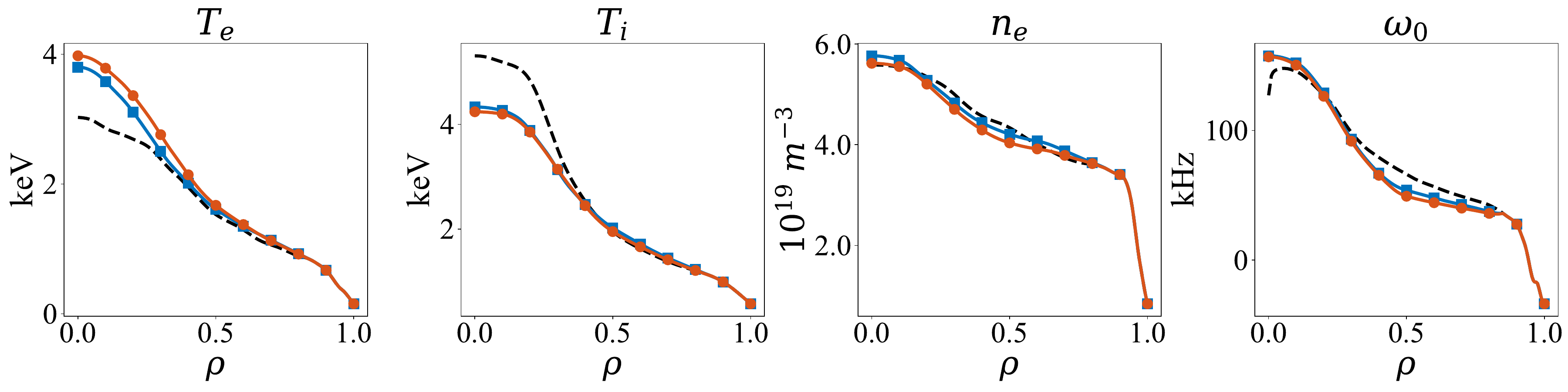}
        \caption{H-mode core profiles}
        \label{fig:profiles_hmode}
    \end{subfigure}
    \caption{\revision{\textbf{Reconstructed core profiles in L-mode and H-mode scenarios}.} Comparison of electron temperature, ion temperature, electron density, and toroidal rotation frequency profiles for experimental measurements, \base, and \ours. (a) L-mode scenario profiles. (b) H-mode scenario profiles. Standalone TGLF does not converge in either scenario and is therefore not shown among the reconstructed profiles.}
    \label{fig:profiles_combined}
\end{figure}



\begin{figure}[!htb]
    \centering
    \begin{subfigure}[b]{0.48\textwidth}
        \centering
        \includegraphics[width=\textwidth]{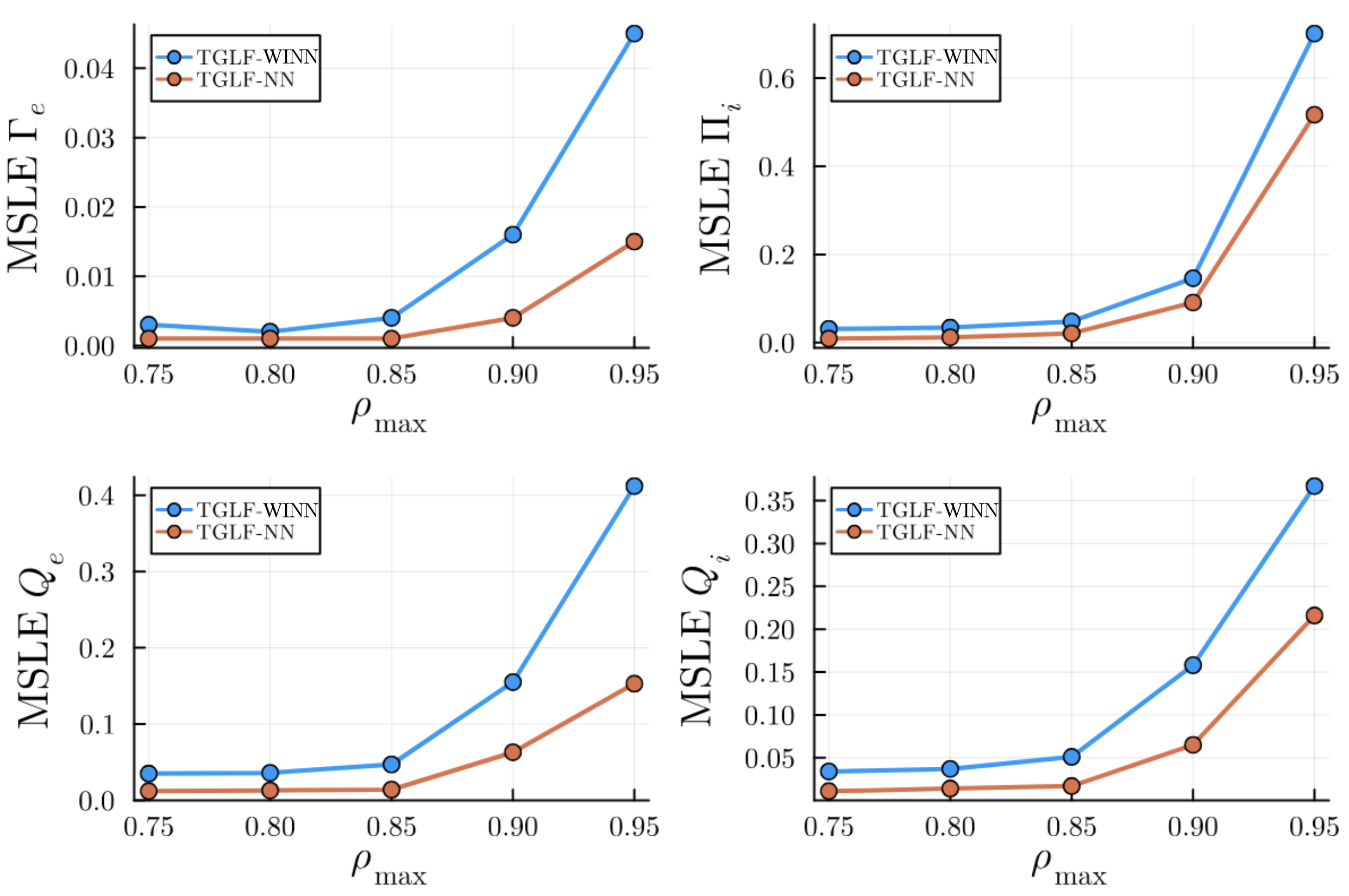}
        \caption{Relative error between the predicted flux and the prescribed source for 3000 DIII-D H-mode plasmas.}
        \label{fig:conv:H}
    \end{subfigure}
    \hfill
    \begin{subfigure}[b]{0.48\textwidth}
        \centering
        \includegraphics[width=\textwidth]{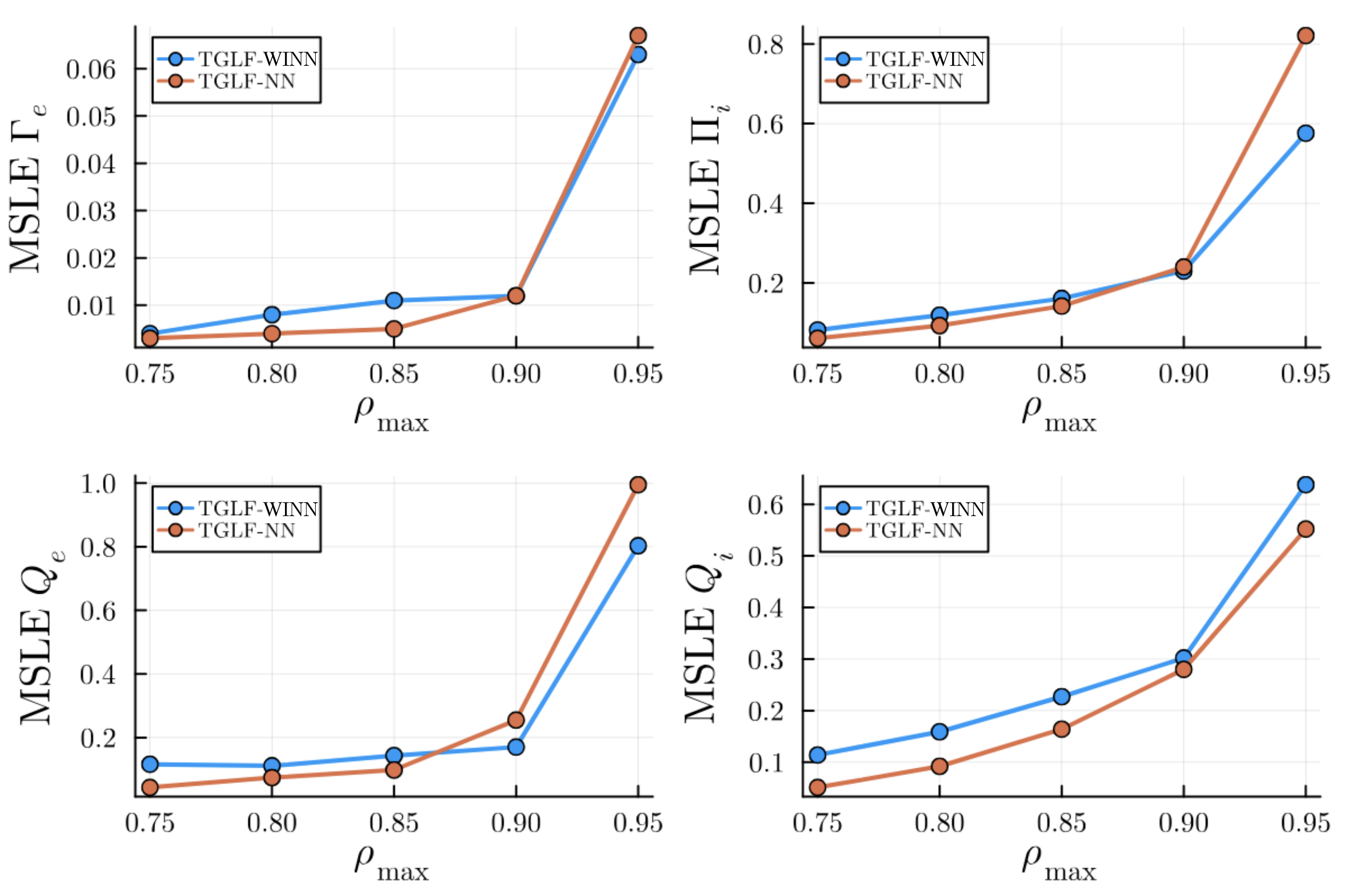}
        \caption{Relative prediction-source error for 1500 DIII-D L-mode plasmas.}
        \label{fig:conv:L}
    \end{subfigure}
    \caption{\revision{\textbf{Comparison of MSLE relative prediction-source errors of \base~and \ours:}} (a) 3000 H-modes and (b) 1500 L-modes. The relative prediction-source errors show a nonlinear increase above $\rho_{\mathrm{max}}=0.85$ for H-modes and above $\rho_{\mathrm{max}}=0.9$ for L-modes, suggesting these as empirical upper limits for the $\rho$-range used for turbulent flux-matching.}
    \label{fig:comparison:convergence}
\end{figure}

\paragraph{Experimental Setup.}
To assess the practicality of \ours, we integrate it into the flux matching workflow of the FUSE integrated modeling suite~\cite{Meneghini24}, which couples the core turbulence, the pedestal (assumed to be fixed as experimental measurements) and the neoclassical transport components. Profiles ($T_e$, $T_i$, $n_e$, $\omega_0$) are reconstructed at 16 uniformly spaced radial locations $\rho \in [0.1, 0.85]$. We compare three core transport options: 1) numerical TGLF solver~\cite{Neiser24a}, 2) \base, and 3) \ours~using two representative DIII-D cases: L-mode discharge (shot \#$161582$ at time $2300$ ms) and H-mode discharge (shot \#$157955$ at time $2700$ ms).

The flux-matching loop enforces steady-state power and particle balance at each radius: the total transport flux will be optimized to equal the target flux set by the prescribed source. With the same four-channel flux vector $\vy = \{\channelone, \channeltwo, \channelthree, \channelfour\}$ used throughout this paper, the source-defined target flux at radius $r$, which we denote $\vy^S$, is
\begin{equation}
    \vy^S(r) = \frac{1}{V'(r)} \int_0^r V'(x)\, \vs(x)\, dx,
    \label{eq:source_target}
\end{equation}
where $\vs(x)$ is the per-channel local source density (auxiliary heating, collisional inter-species exchange, and fueling) and $V'(r) = dV/dr$ is the flux-surface volume element. The ``Source'' curve in Figure~\ref{fig:flux_profiles_combined} plots $\vy^S(r)$, which is known a priori from the experimental deposition profiles.

The coupling is strictly \emph{forward}: given a candidate plasma profile, the turbulent and neoclassical modules each produce flux vectors $\vy^{\rm turb}$ and $\vy^{\rm neo}$ as outputs, and there is no inverse map from a desired flux back to a profile. Steady state is therefore searched for iteratively: the solver adjusts the profile until the forward-evaluated total flux $\vy = \vy^{\rm turb} + \vy^{\rm neo}$ numerically matches $\vy^S$ at every radius. In our setup the NN surrogate (\base~or \ours) or numerical TGLF provides $\vy^{\rm turb}$, while $\vy^{\rm neo}$ is computed by FUSE's standard neoclassical module. We drive this search with FUSE.jl~\cite{Meneghini24}\footnote{\url{https://github.com/ProjectTorreyPines/FUSE.jl}} through the NLsolve.jl backend, using damping coefficient $\beta = -0.25$ and Anderson history depth $m = 5$. Each step applies a Picard update along the negative-residual direction and then corrects it with Type-I Anderson acceleration~\cite{walker2011anderson}. The acceleration maintains a QR-factored history of the last $m$ residual and iterate differences and solves a small least-squares problem whose solution yields a low-rank implicit approximation of the inverse Jacobian.

Because flux intensity times surface area equals the total flux crossing a flux surface, the per-radius residual components are weighted by the flux-surface area normalized to that of the last closed flux surface. Denoting by $y_c(r_i)$ and $y^S_c(r_i)$ the $c$-th channel of $\vy$ and $\vy^S$ evaluated at the $i$-th reconstructed radius $r_i$, the weighted and per-channel-normalized residual is
\begin{equation}
    \tilde{R}_{c,i} = \frac{w_i}{N_c} \left[\, y^S_c(r_i) - y_c(r_i) \,\right], \qquad w_i = \frac{A(r_i)}{A(r_{\rm edge})} \in (0,\,1],
    \label{eq:weighted_residual}
\end{equation}
so that mismatches at outer radii (larger $w_i$) dominate the norm in proportion to the total power they carry. The per-channel normalization $N_c = \tfrac{1}{2}(\|\vy_c \odot \vw\|_2 + \|\vy^S_c \odot \vw\|_2)$, where $\vy_c$ and $\vy^S_c$ stack $y_c(r_i)$ and $y^S_c(r_i)$ over the 16 reconstructed radii and $\vw$ stacks the corresponding weights $w_i$, is set once on the first residual evaluation and renders $\tilde R$ dimensionless. The per-(channel, radius) components $\{\tilde{R}_{c,i}\}$ are assembled into a single residual vector of length $N_\text{channel}\times N_\text{radius}$ that the solver drives toward zero componentwise, with no summation over radii or channels. Convergence is declared once the componentwise maximum reaches the tolerance, $\|\tilde R\|_{\ell^\infty} = \max_{c,i} \lvert \tilde{R}_{c,i} \rvert \leq 10^{-2}$, which reads as ``the worst surface-weighted flux mismatch across all channels and radii is within 1\% of the characteristic flux scale for its channel.'' Any nonzero residual visible in Figure~\ref{fig:flux_profiles_combined} is therefore a search artifact of the iterative procedure at its termination tolerance; exact power balance is recovered in the limit of perfect convergence.

One ``iteration step'' in Figure~\ref{fig:convergence_combined} corresponds to one Picard+Anderson step. Each step invokes one batched surrogate call per radial location, with all 24 $k_y$ branches evaluated in parallel within a call. The total number of surrogate forward passes per run is therefore the reported iteration count multiplied by the number of reconstructed radii (16 in our setup).

\paragraph{Convergence Performance.}
Both NN surrogates exhibit faster and more stable convergence than the numerical solver. As shown in Figure~\ref{fig:convergence_combined}, \ours~converges in 129 iterations for the L-mode and 412 iterations for H-mode, while \base~converges in 251 and 364 iterations, respectively. The numerical solver fails to converge after 1000 iterations in both scenarios.

Examining individual transport channels reveals that the numerical solver exhibits poor convergence in the $\omega_0$ channel for the L-mode and struggles across all channels in the H-mode. We attribute the superior NN performance to the fact that TGLF produces point-wise noisy fluxes across neighboring radii, most visibly as a $10$--$15\times$ spike near $\rho \approx 0.3$ in H-mode that is consistent with the Hermite-eigensolver truncation/convergence issues already acknowledged in Section~\ref{sec:dataset} (see Appendix~\ref{sec:appendix_tglf_at_nn} for a direct visualization of these point-wise TGLF flux oscillations at the WINN-converged state), and an Anderson-accelerated flux-matching solver cannot zero a residual that is not Lipschitz-smooth in the profile variables. The NN surrogates present a smooth residual surface to the solver and therefore converge stably.

\paragraph{Reconstruction Quality.}
As shown in Figure~\ref{fig:flux_profiles_combined}, \ours~and \base~consistently produce lower final mismatch values between predicted and prescribed source transport fluxes in both L-mode and H-mode scenarios, while TGLF exhibits significantly higher flux mismatches. The reconstructed core profiles (Figure~\ref{fig:profiles_combined}) show good agreement with the experimental measurements across all channels, with deviations observed in low $\rho$ regions (near the magnetic axis) for $T_e$ in L-mode and both $T_e$ and $T_i$ in H-mode. This near-axis mismatch reflects two compounding effects, neither attributable to the NN surrogate itself. First, experimental profile diagnostics are intrinsically less precise close to the magnetic axis: Thomson scattering and charge-exchange recombination spectroscopy have reduced spatial resolution and poorer photon statistics at small minor-radius volumes, and line-integrated diagnostics are geometrically insensitive to the small-volume core region; the ``experimental'' curves at $\rho \lesssim 0.2$ therefore carry larger uncertainty than at mid-radius. Second, TGLF's local quasi-linear formulation becomes progressively less reliable near axis: the ratio of ion sound gyroradius to minor radius $\rho_s/a$ grows, finite-banana-width effects become non-negligible, and non-local transport physics such as sawtooth reconnection is outside the scope of a standalone quasilinear model. Because our surrogate is trained on TGLF outputs, it inherits these limitations of the underlying ROM; the near-axis deviation is therefore a bound imposed by the physics model rather than a deficiency of the NN approximation.

\paragraph{Computational Efficiency Gain.}
Due to the faster inference time and fewer iterations required for convergence, the flux-matching procedure with \ours~takes 20 seconds compared to 15 minutes with the numerical solver using 128 threads, representing a 45× speed-up. \base~achieves a 600× speed-up (1.5 seconds), likely due to its native Julia implementation being more optimized for this workflow. Nevertheless, the 45× speed-up of \ours~demonstrates substantial efficiency advantages, and its translation into a Julia implementation could enable further speed-up.

\begin{figure}[!h]
    \centering
    \includegraphics[width=0.5\textwidth]{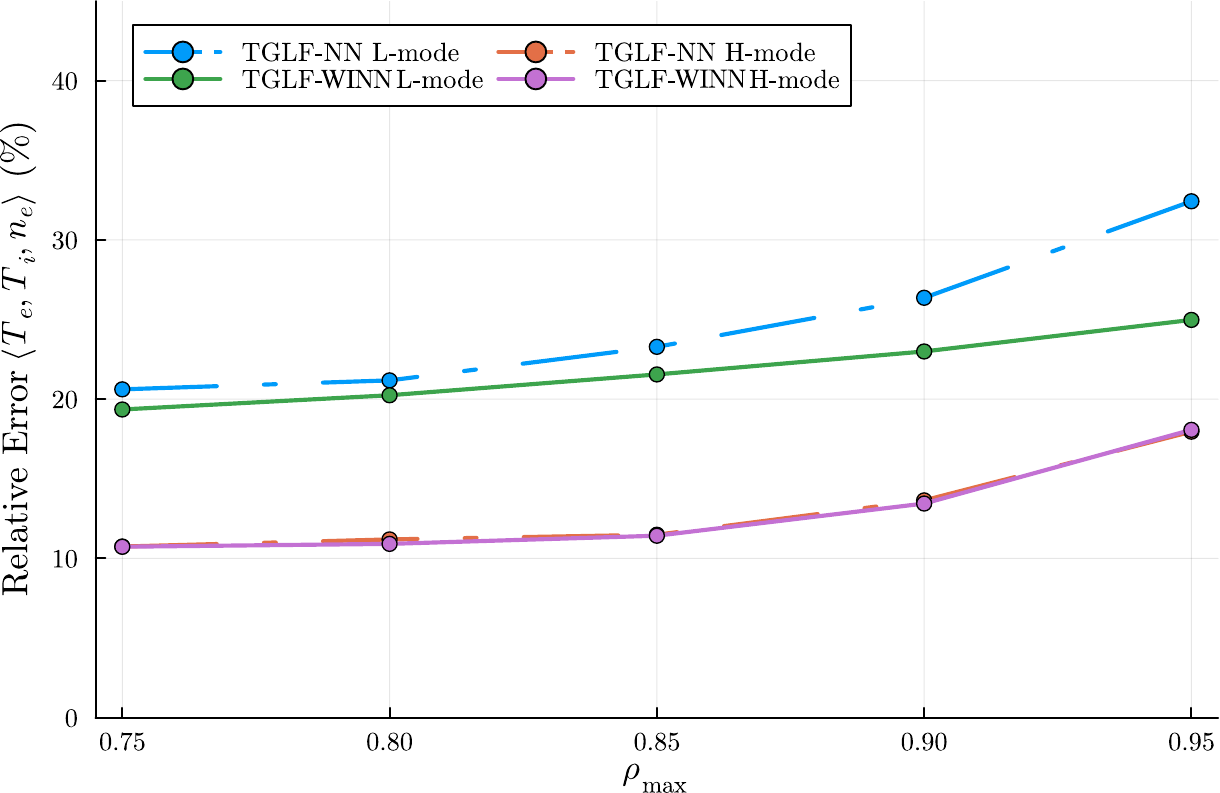}
    \caption{\revision{\textbf{Per-radius relative error comparison}.} Mean absolute relative error for reconstructed profiles ($T_e$, $T_i$, $n_e$) as a function of $\rho_{\mathrm{max}}$ in H-mode and L-mode across the DIII-D validation dataset.}
    \label{fig:rhomax_comp}
\end{figure}

\paragraph{Extended Dataset Validation.}
We extended validation to 3000 H-mode and 1500 L-mode DIII-D profiles, following large-database validation methodologies established in the transport modeling community~\cite{abbate2024large}. A lumped confinement sanity check of both NN surrogates against the IPB98(y,2) scaling law, based on the volume-integrated stored thermal energy $W_{\rm th}$ (Equation~\ref{eq:Wth}), is reported in Appendix~\ref{sec:appendix:wth_lumped}.

The per-radius relative errors (Figure~\ref{fig:rhomax_comp}) show that \ours~delivers significant performance improvements in L-mode while H-mode performance remains comparable to \base. A complementary aggregated mean-and-standard-deviation view of predicted profiles across the full 3000-H-mode / 1500-L-mode validation dataset is provided in Appendix~\ref{sec:appendix:aggregated_validation}. On the relative error between the predicted flux and the prescribed source (Figure~\ref{fig:comparison:convergence}), \base~consistently achieves smaller values, likely due to ensemble averaging (20 networks) versus our single network.

\section{Conclusion and Future Work}
We present \ours, a data-efficient neural surrogate for TGLF that matches \base's full-data offline accuracy using only 25\% of the training data and delivers a 12.5\% relative RMSLE reduction when both are trained on the full dataset. The method additionally exhibits an order-of-magnitude smaller performance degradation than \base~under sparse, unfiltered training, attributable to the wavenumber-informed regularization providing a physics-guided constraint on per-mode flux contributions.

\paragraph{Limitations.}
While our wavenumber-conditioned architecture leverages physics structure through wavenumber decomposition, the current implementation does not explicitly enforce fundamental conservation laws (energy, momentum, particle conservation) as hard constraints in the neural network architecture. Although the training targets from TGLF satisfy these physical constraints, the surrogate predictions may occasionally violate them due to approximation errors. Additionally, systematic biases observed in flux-matching applications stem from underlying TGLF ROM limitations in certain plasma regimes (e.g., plasma edge regions with large $\rho$), where TGLF's quasi-linear assumptions break down.

Addressing these limitations and extending the core framework naturally suggests several directions for future work. Several promising extensions are worth exploring. First, combining our approach with Mixture-of-Experts frameworks for explicit mode classification (ITG vs. ETG) could further concentrate flux distributions and improve specialized mode predictions. Second, the per-branch parity analysis in Appendix~\ref{sec:appendix_per_branch_parity} shows that our network's branch-level outputs are individually accurate across the ITG, TEM, and ETG wavenumber regimes; combined with TGLF's linear-stability diagnostics (growth rate, real frequency, critical-gradient indicators), these per-branch outputs could enable a data-driven identification of the dominant instability mode at each plasma condition. Third, inter-branch connections in the per-wavenumber network could further exploit cross-scale coupling present in the SAT2/SAT3 saturation rules. Fourth, physics-informed neural network (PINN) techniques~\cite{raissi2019physics} or constrained optimization layers could be incorporated to guarantee hard conservation-law consistency in the surrogate predictions. Fifth, learning linear mode solutions directly which can analytically calculate the quasilinear weights and intensities in a more decomposed fashion, enabling zero-shot generation to new saturation rules without retraining. Sixth, \ours~per-wavenumber predictions could serve as synthetic diagnostics for between-shot predictions of theoretically interesting wavenumber ranges for Doppler Back-Scattering diagnostics~\cite{Pratt24}. Finally, our TGLF data and active learning approach could provide warm starts for surrogating higher-fidelity gyrokinetic models such as CGYRO~\cite{candy2016high,Neiser23}: at CGYRO's cost of hours of supercomputer time per sample, the $4\times$ data reduction demonstrated here would translate into thousands of GPU-hours saved per trained surrogate.

\section*{Acknowledgments}
This material is based upon work supported by the U.S. Department of Energy, Office of Science, Office of Fusion Energy Sciences, using the DIII-D National Fusion Facility, a DOE Office of Science user facility, under Awards DE-FC02-04ER54698 (DIII-D), DE-SC0024426 (FDP), DE-SC0018990 (MAST-U), DE-SC0022031 (ALF), DE-SC0017992 (AToM), DE-FG02-95ER54309 (GA Theory), and the DOE Office of Workforce Development for Teachers and Scientists (WDTS) under the Science Undergraduate Laboratory Internships (SULI) program. This work used computational resources of the National Energy Research Scientific Computing Center (NERSC), a Department of Energy User Facility using NERSC awards FES-ERCAP 30971 and FES-ERCAP 33303, and the computational resources of NRP Nautilus, which is supported by National Science Foundation (NSF) awards CNS-1730158, ACI-1540112, ACI-1541349, OAC-1826967, OAC-2112167, CNS-2100237, CNS-2120019.

\section*{Disclaimer}
This report was prepared as an account of work sponsored by an agency of the United States Government. Neither the United States Government nor any agency thereof, nor any of their employees, makes any warranty, express or implied, or assumes any legal liability or responsibility for the accuracy, completeness, or usefulness of any information, apparatus, product, or process disclosed, or represents that its use would not infringe privately owned rights. Reference herein to any specific commercial product, process, or service by trade name, trademark, manufacturer, or otherwise, does not necessarily constitute or imply its endorsement, recommendation, or favoring by the United States Government or any agency thereof. The views and opinions of authors expressed herein do not necessarily state or reflect those of the United States Government or any agency thereof.

\newpage
\bibliographystyle{unsrt}
\bibliography{references}

\newpage
\appendix
\onecolumn

\section{Complete Ablation Results}
\label{sec:appendix_ablation}

This section presents comprehensive offline training results for all combinations of contributions (\base, \base~+ \firstcontri, \base~+ \secondcontri, \ours) across both datasets (\MaMiPert, \MaPert), with and without filtering. We evaluated the performance of the model using the root mean squared logarithmic error (RMSLE, Equation \ref{eq:rmsle}) for each flux channel and the average $R^2$ in all channels.

The $R^2$ metric is defined in Equation (\ref{eq:r2}), where \(N\) is the number of test samples, \(\vy'_i\) represents the predicted fluxes, \(\vy_i\) represents the TGLF target fluxes and \(\bar{\vy}\) represents the mean of the TGLF target fluxes on the test set.

\begin{equation}
    R^2(\vy', \vy) = 1 - \frac{\sum_{i=1}^{N} \left( \vy_i - \vy'_i \right)^2}{\sum_{i=1}^{N} \left( \vy_i - \bar{\vy} \right)^2},
    \label{eq:r2}
\end{equation}

\begin{table}[!h]
    \centering
    \caption{\textbf{Comprehensive ablation results across all experimental configurations.} Performance metrics are reported as Root Mean Squared Logarithmic Error (RMSLE) ($\times 10^{-2}$) on test sets, except for the final column which shows $R^2$ values. Best results are highlighted in \textbf{bold}. Abbreviations: \firstcontri~= Feature Tuning, \secondcontri~= Wavenumber Regularization, Ma. = \MaPert, Mi. = Minor Perturbation, F = Unfiltered, T = Filtered.}
    \label{tab:ablation_results}
    \resizebox{0.9\textwidth}{!}{
        \centering
        \begin{tabular}{lcccccccc}
            \toprule
            \textbf{Model}        & \textbf{Dataset} & \textbf{Filtered} & \(\channelone\)        & \(\channeltwo\)        & \(\channelthree\)      & \(\channelfour\)       & \textbf{Avg.}          & \textbf{$R^2$}  \\
            \midrule
            \base~                & Ma.              & F                 & $24.41\pm5.67$         & $41.8\pm4.18$          & $51.96\pm5.69$         & $36.55\pm4.91$         & $39.93\pm4.72$         & -8.80E+11       \\
            \base~                & Ma.              & T                 & $6.28\pm1.75$          & $9.12\pm1.46$          & $12.06\pm2.35$         & $8.82\pm2.06$          & $9.3\pm1.45$           & 0.9221          \\
            \base~                & Ma. + Mi.        & F                 & $32.2\pm9.62$          & $47.74\pm10.66$        & $57.03\pm9.62$         & $44.69\pm10.62$        & $46.28\pm10.54$        & -1.84E+19       \\
            \base~                & Ma. + Mi.        & T                 & $4.23\pm1.08$          & $6.54\pm0.94$          & $8.76\pm1.6$           & $6.33\pm1.41$          & $6.66\pm0.9$           & \textbf{0.9685} \\
            \midrule
            \base~+ \firstcontri  & Ma.              & F                 & $10.75\pm1.99$         & $13.36\pm2.43$         & $15.46\pm2.8$          & $13.24\pm2.08$         & $ 13.31\pm2$           & 0.1022          \\
            \base~+ \firstcontri  & Ma.              & T                 & $6.02\pm1.66$          & $8.54\pm1.5$           & $11.38\pm2.31$         & $8.11\pm1.95$          & $8.73\pm1.42$          & 0.9178          \\
            \base~+ \firstcontri  & Ma. + Mi.        & F                 & $7.71\pm1.3$           & $9.64\pm1.62$          & $11.49\pm2.18$         & $9.59\pm1.46$          & $ 9.7\pm1.46$          & 0.4343          \\
            \base~+ \firstcontri  & Ma. + Mi.        & T                 & $3.9\pm0.98$           & $5.8\pm0.97$           & $7.86\pm1.61$          & $5.49\pm1.22$          & $5.93\pm0.87$          & 0.9646          \\
            \midrule
            \base~+ \secondcontri & Ma.              & F                 & $21.51\pm3.64$         & $28.71\pm3.73$         & $31.34\pm4.01$         & $27.5\pm3.54$          & $27.5\pm3.33$          & -4.62E+07       \\
            \base~+ \secondcontri & Ma.              & T                 & $6.22\pm1.69$          & $8.96\pm1.51$          & $11.82\pm2.37$         & $8.68\pm2.03$          & $9.14\pm1.43$          & 0.9147          \\
            \base~+ \secondcontri & Ma. + Mi.        & F                 & $21.59\pm3.63$         & $28.54\pm3.33$         & $30.8\pm3.48$          & $26.73\pm3.14$         & $27.13\pm3.04$         & -1.37E+11       \\
            \base~+ \secondcontri & Ma. + Mi.        & T                 & $4.16\pm1.08$          & $6.25\pm0.89$          & $8.42\pm1.58$          & $ 6\pm1.32$            & $6.39\pm0.87$          & 0.9679          \\
            \midrule
            \ours                 & Ma.              & F                 & $9.97\pm2$             & $12.92\pm2.97$         & $14.53\pm2.93$         & $12.43\pm2.21$         & $12.57\pm2.18$         & 0.3760          \\
            \ours                 & Ma.              & T                 & $5.93\pm1.71$          & $8.47\pm1.56$          & $11.4\pm2.47$          & $7.92\pm2.03$          & $8.66\pm1.5$           & 0.9016          \\
            \ours                 & Ma. + Mi.        & F                 & $6.62\pm1.32$          & $8.33\pm1.68$          & $9.98\pm2.09$          & $8.18\pm1.49$          & $8.36\pm1.43$          & 0.6482          \\
            \ours                 & Ma. + Mi.        & T                 & \textbf{3.85$\pm$0.99} & \textbf{5.68$\pm$0.92} & \textbf{7.75$\pm$1.55} & \textbf{5.35$\pm$1.18} & \textbf{5.83$\pm$0.84} & 0.9639          \\
            \bottomrule
        \end{tabular}}
\end{table}

\section{Hyperparameter Search Results}
\label{sec:appendix:hyperparam}

\begin{figure}[!h]
    \centering
    \includegraphics[width=0.6\textwidth]{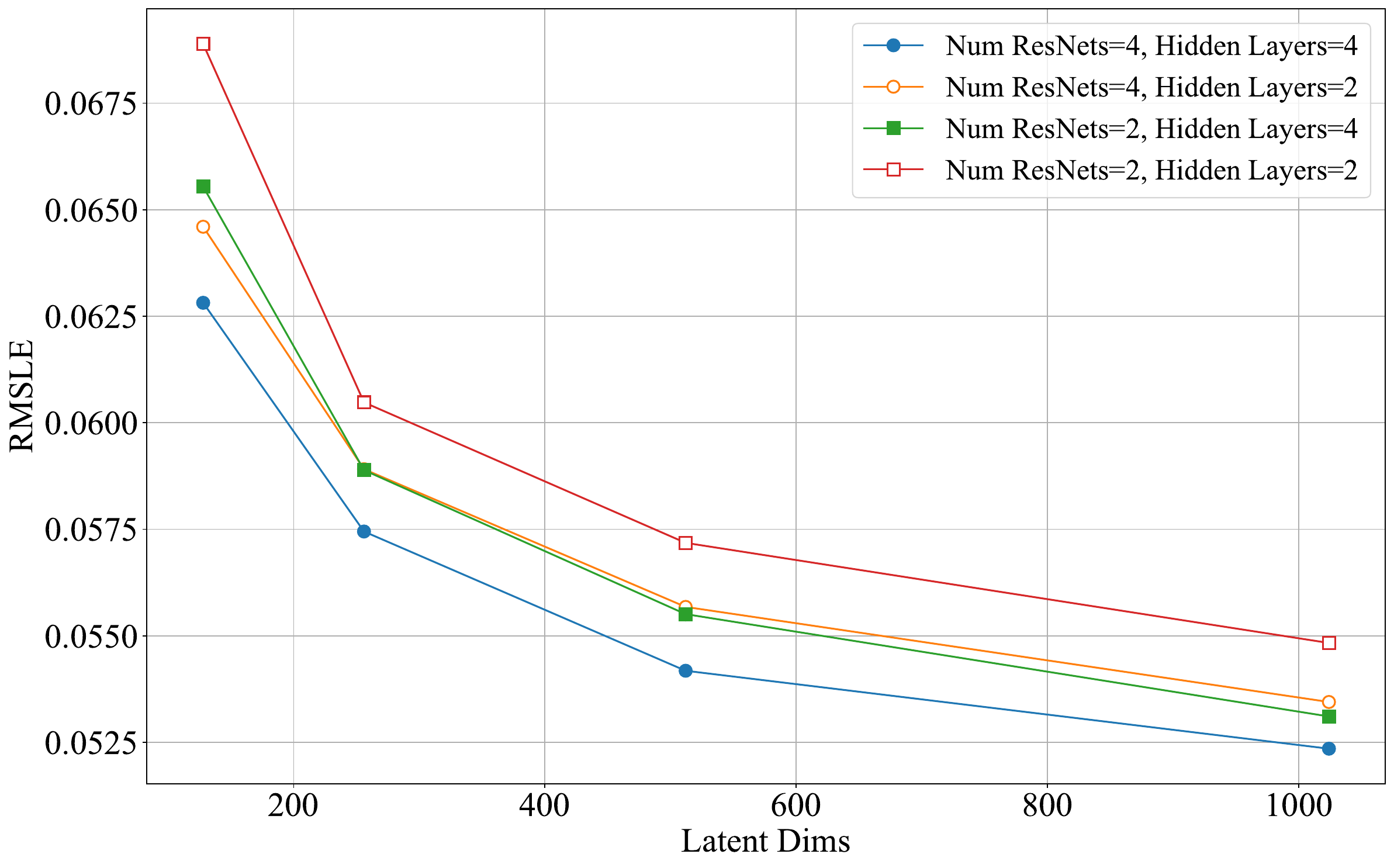}
    \caption{Impact of latent dimensions on model performance. The graph shows the test RMSLE of \ours~on the filtered \MaMiPert~dataset as a function of the number of latent dimensions, with all other hyperparameters held constant. Lower RMSLE values indicate superior performance.}
    \label{fig:hyperparam:latent_dim}
\end{figure}

\begin{figure}[!h]
    \centering
    \includegraphics[width=0.6\textwidth]{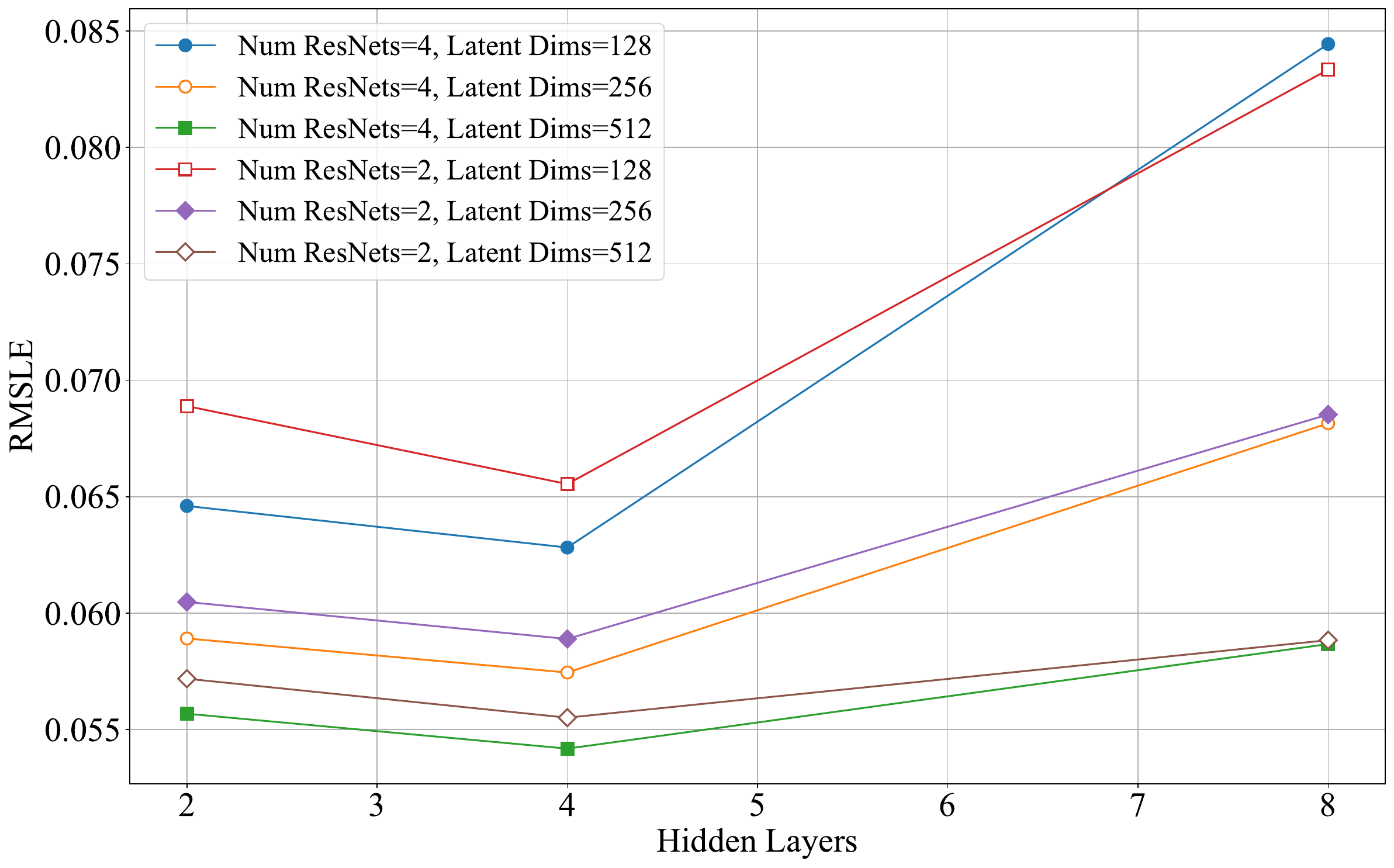}
    \caption{Effect of hidden layer depth on model performance. The graph illustrates the test RMSLE of \ours~on the filtered \MaMiPert~dataset as a function of the number of hidden layers, with all other hyperparameters held constant. Lower RMSLE values indicate superior performance.}
    \label{fig:hyperparam:hidden_layer}
\end{figure}

\begin{figure}[!h]
    \centering
    \includegraphics[width=0.6\textwidth]{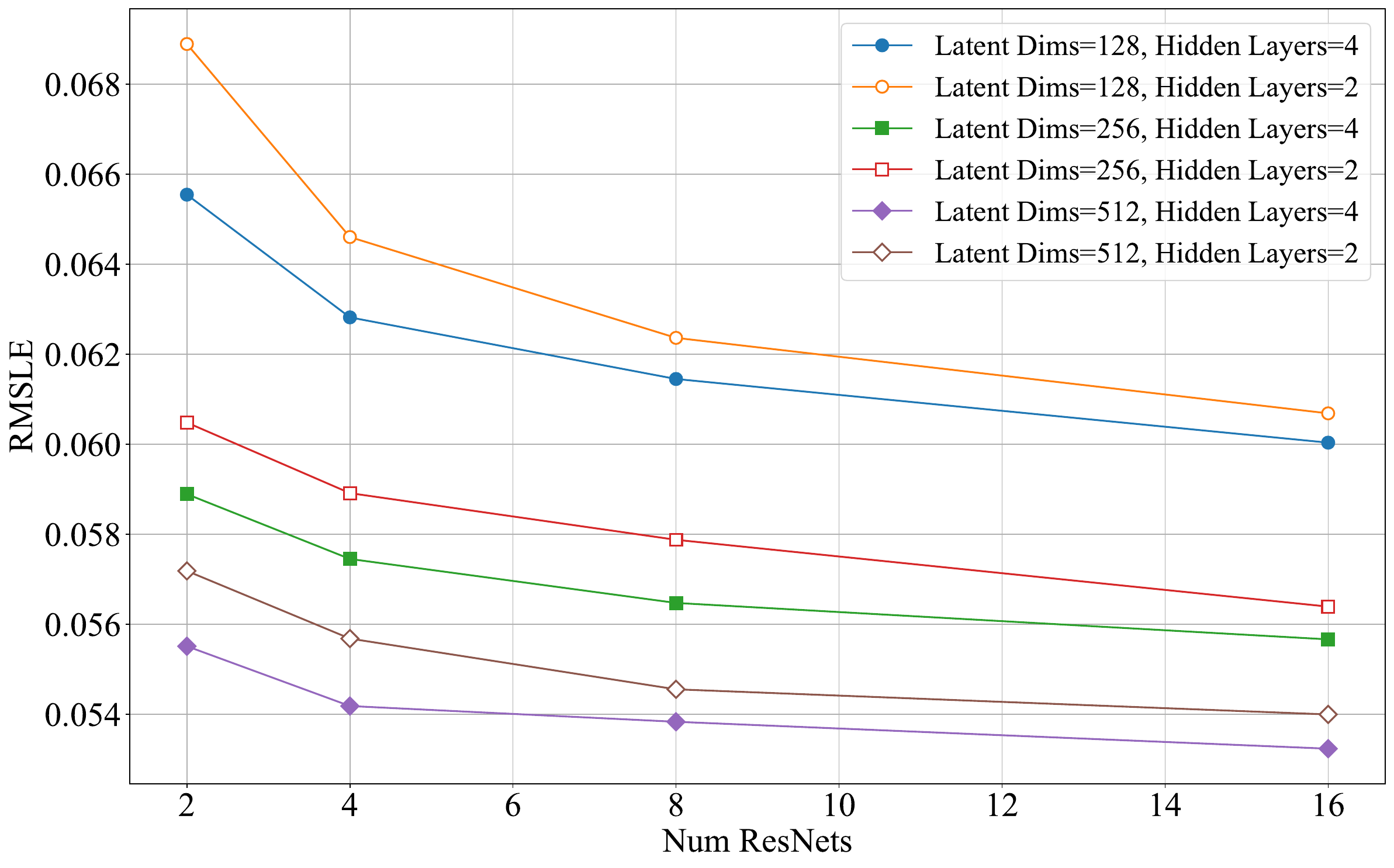}
    \caption{Influence of ResNet quantity on model performance. The graph depicts the test RMSLE of \ours~on the filtered \MaMiPert~dataset as a function of the number of ResNets, with all other hyperparameters held constant. Lower RMSLE values indicate superior performance.}
    \label{fig:hyperparam:num_resnet}
\end{figure}

We conducted extensive hyperparameter searches for \ours~using the filtered \MaMiPert~dataset. We noted diminishing marginal benefits when the latent dimension exceeded 256 and the number of ResNet blocks exceeded 4 (Figures~\ref{fig:hyperparam:latent_dim} and \ref{fig:hyperparam:num_resnet}). We also observed that the minimal loss was achieved with 4 hidden layers in the MLP (Figure~\ref{fig:hyperparam:hidden_layer}). Collectively, these results guided our choice of hyperparameters in the implementation, balancing accuracy and computational cost. We choose 4 total ResNet blocks, 4 hidden layers per ResNet block, and 256 neurons per hidden layer. These hyperparameters remain consistent for all L-models, including \base, as to eliminate the variable of parameter count.

\section{BAL Dataset Distribution Analysis}
\label{sec:appendix_bal_dist}

\begin{figure}[!h]
    \centering
    \includegraphics[width=0.95\textwidth]{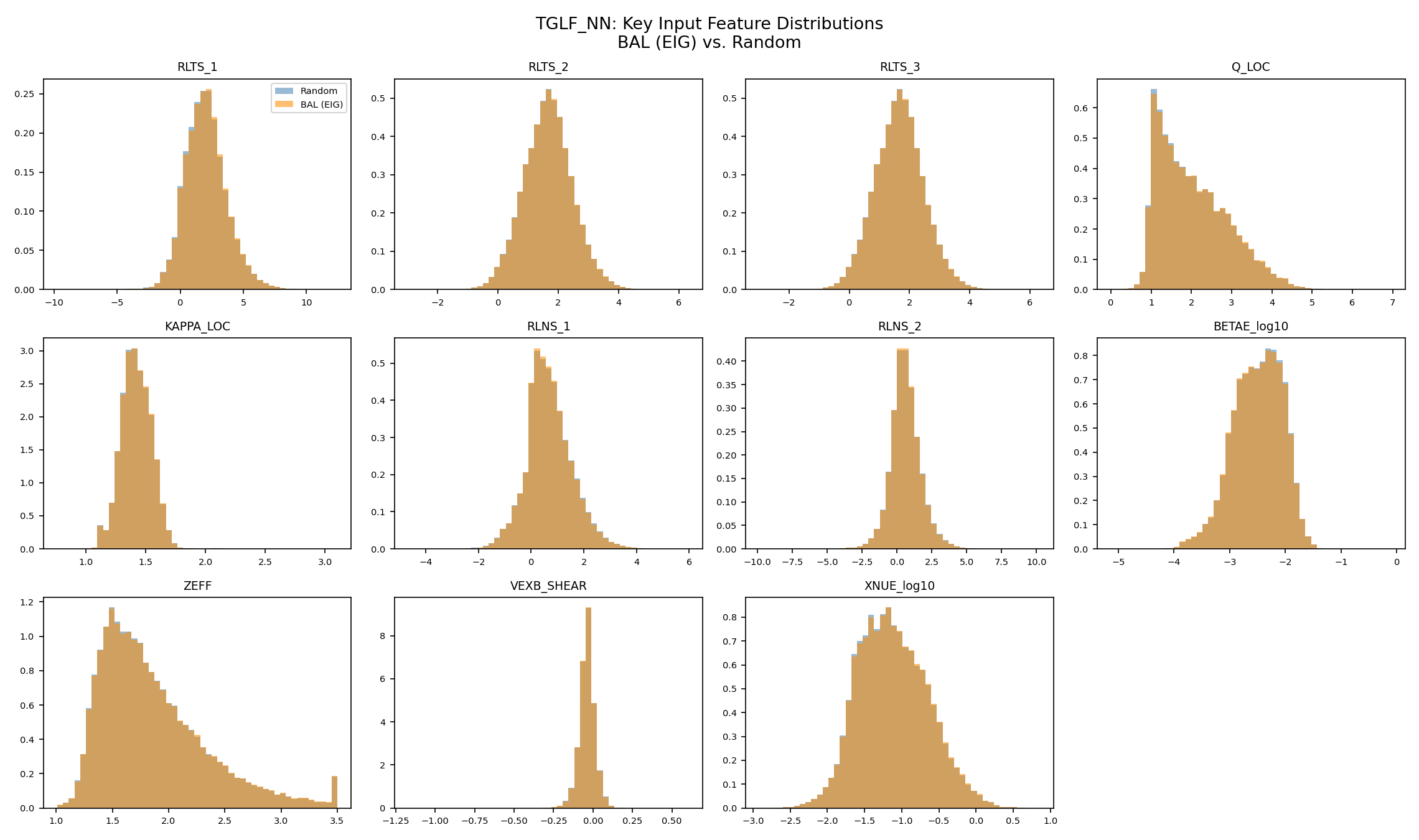}
    \caption{\textbf{\base: key input-feature distributions, BAL (EIG) vs. random.} Overlapping histograms for 11 representative inputs across $\sim\!10^5$ acquired samples per strategy.}
    \label{fig:bal_dist_tglfnn_key}
\end{figure}

\begin{figure}[!h]
    \centering
    \includegraphics[width=0.95\textwidth]{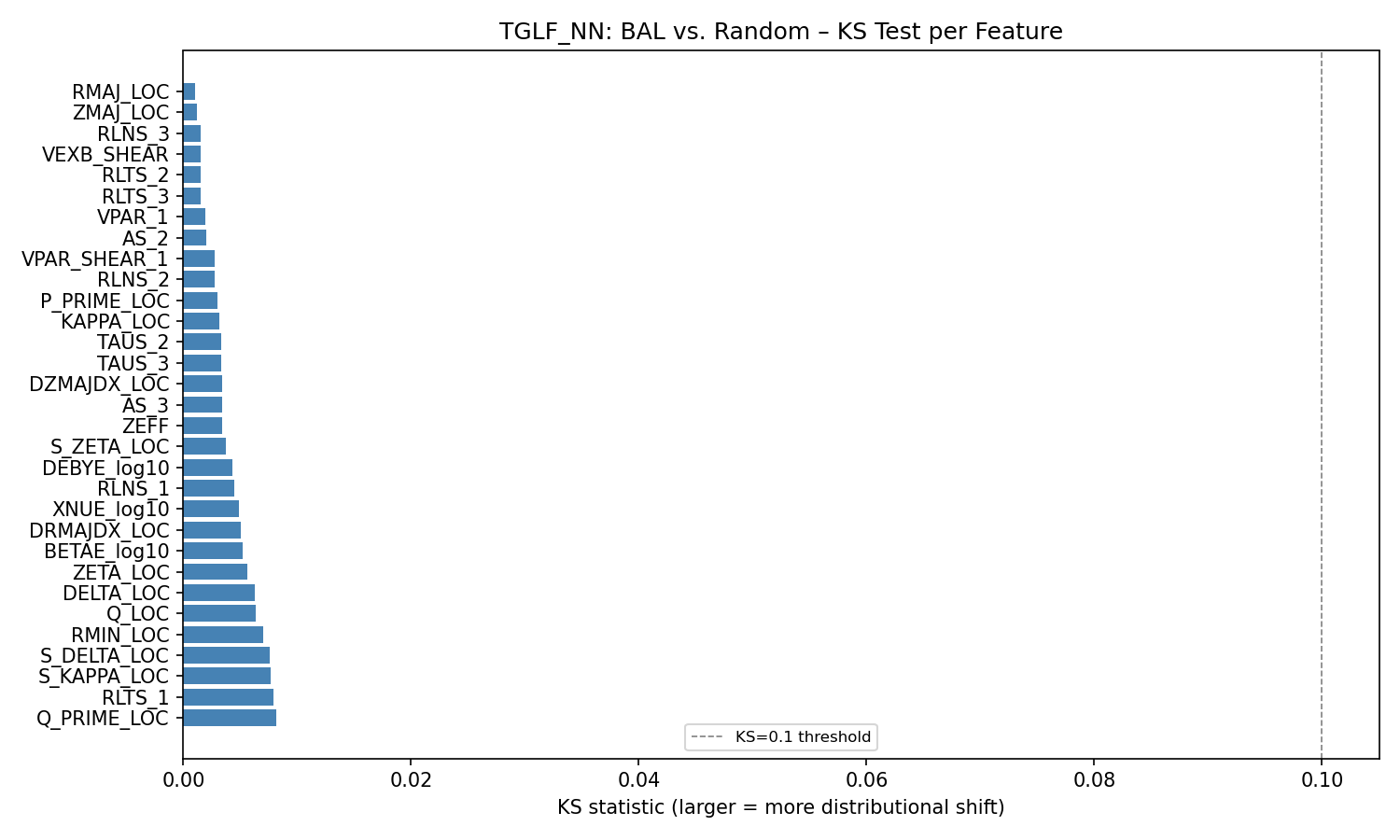}
    \caption{\textbf{\base: per-feature Kolmogorov--Smirnov statistics, BAL vs. random.} All features fall well below the KS$=0.1$ threshold (dashed gray line) for distinguishable distributions.}
    \label{fig:bal_dist_tglfnn_ks}
\end{figure}

\begin{figure}[!h]
    \centering
    \includegraphics[width=0.95\textwidth]{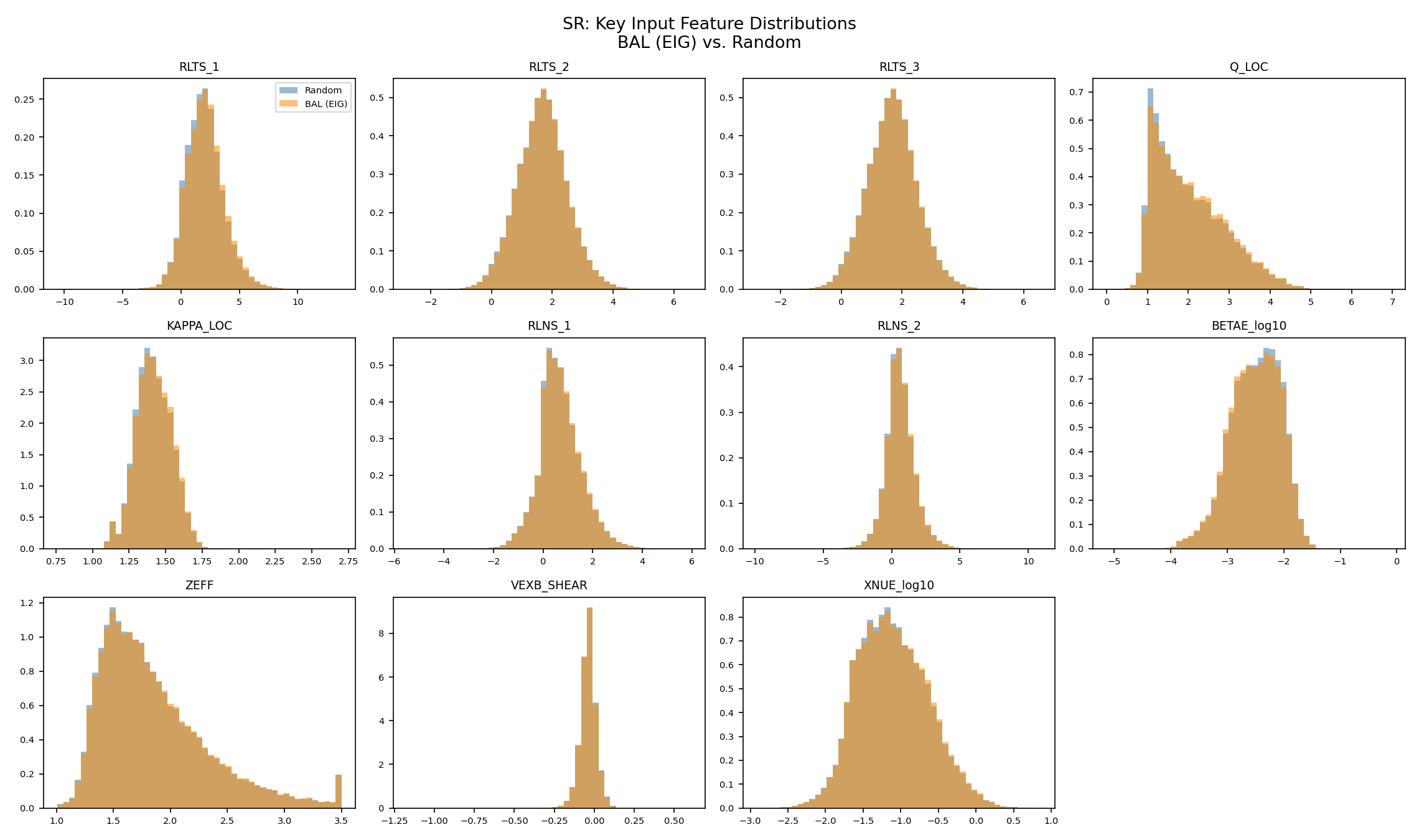}
    \caption{\textbf{\ours: key input-feature distributions, BAL (EIG) vs. random.} Same layout as Figure~\ref{fig:bal_dist_tglfnn_key}.}
    \label{fig:bal_dist_winn_key}
\end{figure}

\begin{figure}[!h]
    \centering
    \includegraphics[width=0.95\textwidth]{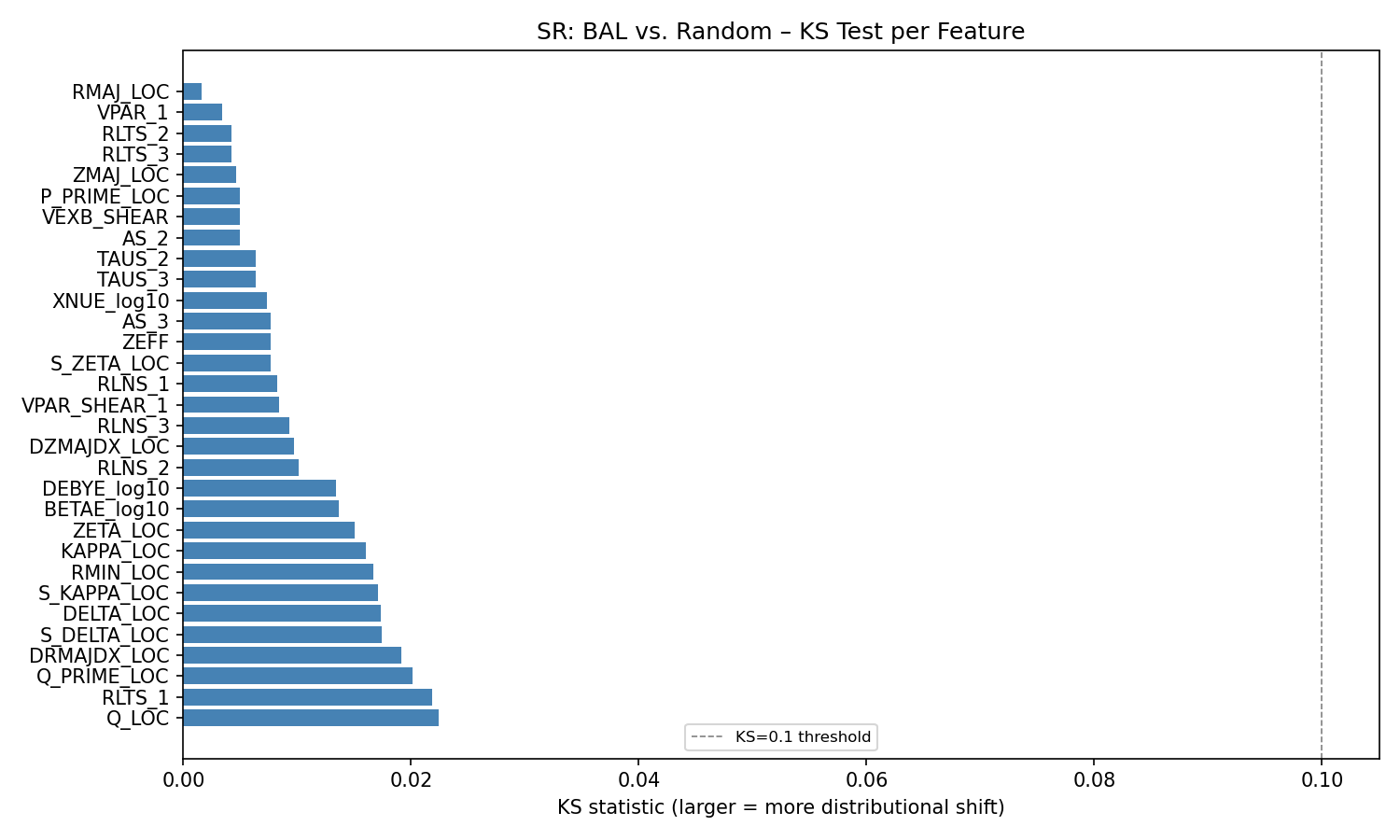}
    \caption{\textbf{\ours: per-feature Kolmogorov--Smirnov statistics, BAL vs. random.} All features fall well below the KS$=0.1$ threshold.}
    \label{fig:bal_dist_winn_ks}
\end{figure}

This appendix supplements Section~\ref{sec:res:bal} with the input-feature distribution comparison between BAL-acquired (EIG) and randomly-acquired training pools, motivated by the Adjudicator's suggestion (following~\cite{ho2025efficient}) to distinguish BAL contributions beyond final model performance. For both \base~and \ours, we pool all BAL cycles ($\sim\!10^5$ acquired samples) and compare against an equal-size random-acquired pool. Key-feature histograms (Figures~\ref{fig:bal_dist_tglfnn_key}, \ref{fig:bal_dist_winn_key}) overlap nearly perfectly, and per-feature Kolmogorov--Smirnov statistics (Figures~\ref{fig:bal_dist_tglfnn_ks}, \ref{fig:bal_dist_winn_ks}) remain below $0.01$ for every input in \base~and below $0.025$ for every input in \ours, well under the commonly used KS$=0.1$ threshold for distinguishable distributions. The two pools are therefore statistically nearly identical in their input marginals.

We interpret this as evidence that the measurable BAL advantage at 25\% data utilization (Figure~\ref{fig:bal_compare_acq_function}) does not come from a gross redistribution of input space. Several factors plausibly contribute, in order of importance: (i) EIG selects on model-output uncertainty, a nonlinear function of inputs whose effect on joint or correlational structure need not manifest in 1D input marginals; (ii) the per-radius candidate-proposing prior (Section~\ref{sec:method:BAL}) imposes the bulk of the structural bias shared by both strategies, further compressing any marginal gap between EIG and random; and (iii) our pool-based BAL draws from the same underlying dataset as random; in the large-acquisition limit the two pools therefore converge in distribution, consistent with observations in~\cite{ho2025efficient}. The acquisition function's advantage therefore arises from sample-level ordering within a distributionally similar pool rather than from restructuring the input space.

\section{TGLF Evaluated at WINN-Converged Profiles}
\label{sec:appendix_tglf_at_nn}

This appendix supports the ``Reconstruction Quality'' discussion in Section~\ref{sec:res:fluxmatching}. For both representative DIII-D cases, we first converge the flux-matching loop using \ours~as the core-transport surrogate, then (i) evaluate TGLF at the WINN-converged profile to obtain point-wise TGLF fluxes at that state, and (ii) warm-start TGLF from the WINN-converged profile and run one additional flux-matching step to see where TGLF ``wants to move.''

\begin{figure}[!h]
    \centering
    \includegraphics[width=1.0\textwidth]{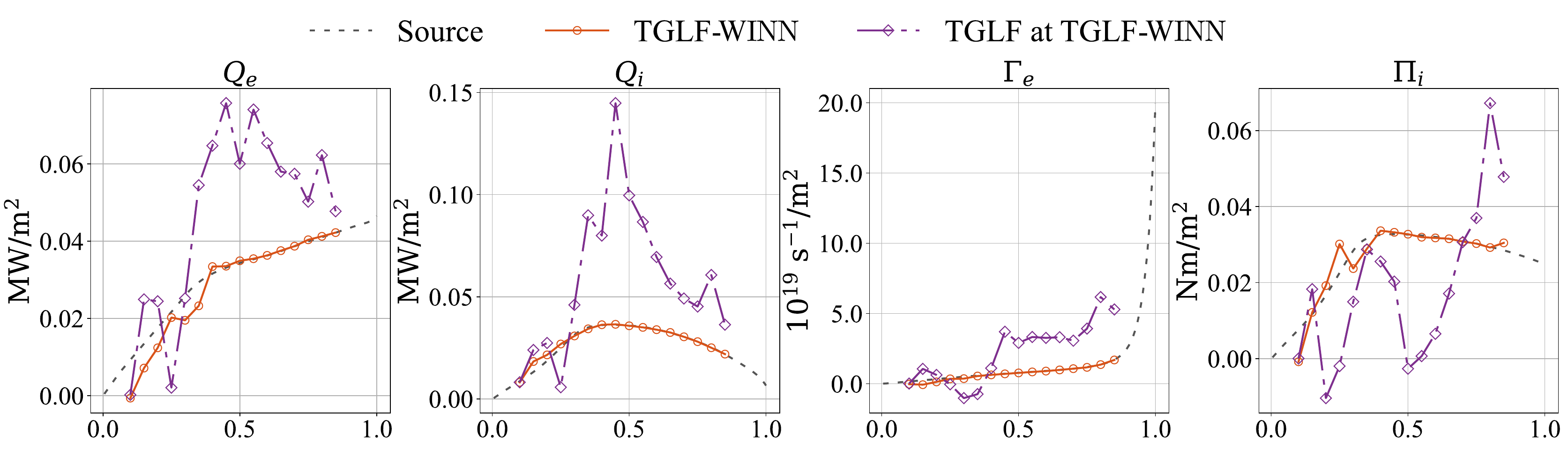}
    \caption{\revision{\textbf{L-mode: TGLF fluxes evaluated at WINN's converged profile.}} WINN's own fluxes track the prescribed Source smoothly; TGLF at the same profile produces $2$--$3\times$ oscillations across neighboring radii in all four channels, a signature of numerical noise in the eigensolver rather than physical discontinuity. This pointwise noise is the root cause of standalone TGLF non-convergence in Figure~\ref{fig:convergence_combined}.}
    \label{fig:tglf_at_nn_L}
\end{figure}

\begin{figure}[!h]
    \centering
    \includegraphics[width=1.0\textwidth]{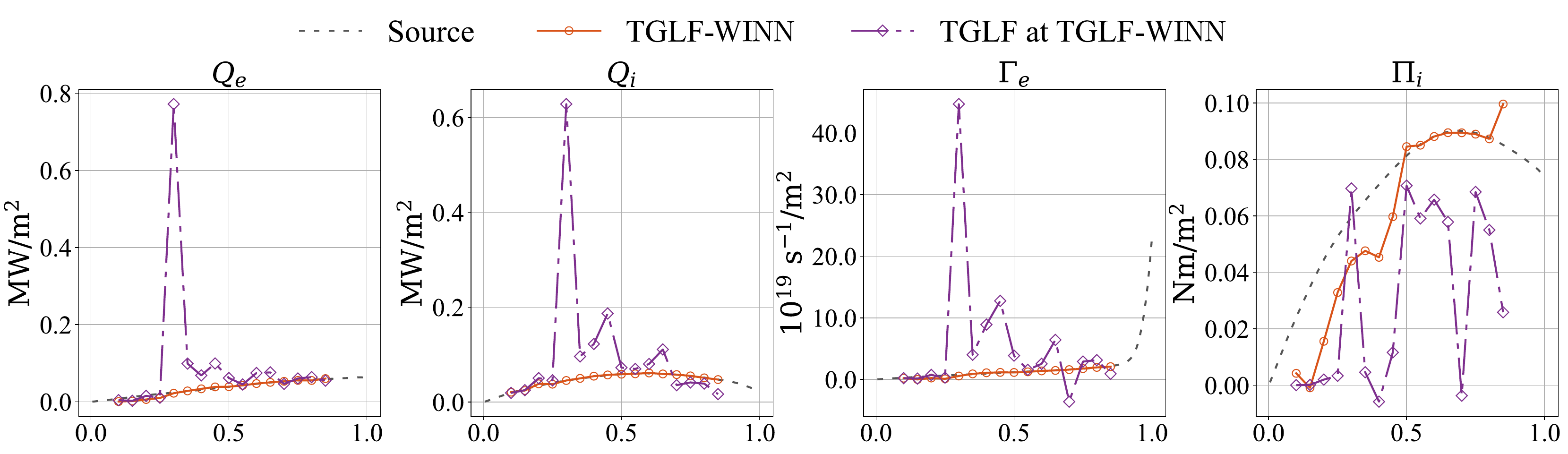}
    \caption{\revision{\textbf{H-mode: TGLF fluxes evaluated at WINN's converged profile.}} A $10$--$15\times$ flux spike appears near $\rho \approx 0.3$ across $Q_e$, $Q_i$, and $\Gamma_e$ at a single radial point, consistent with the Hermite-eigensolver truncation/convergence issues acknowledged in Section~\ref{sec:dataset}. WINN's own predictions remain smooth through this region.}
    \label{fig:tglf_at_nn_H}
\end{figure}

First, TGLF's fluxes at WINN's converged profile are highly oscillatory and do not match the prescribed source (Figures~\ref{fig:tglf_at_nn_L}, \ref{fig:tglf_at_nn_H}), despite WINN's own predictions tracking the source closely; the oscillations are most pronounced near $\rho \approx 0.3$ in H-mode, a pattern consistent with the Hermite-eigensolver truncation/convergence issues acknowledged in Section~\ref{sec:dataset} and indicating numerical noise across neighboring radial points rather than a physical discontinuity.

Second, as a consequence of this noise pattern, the Anderson-accelerated flux-matching iteration, which requires a Lipschitz-smooth residual, is fundamentally obstructed by the pointwise noise rather than by a suboptimal optimization trajectory, explaining standalone TGLF non-convergence in Figure~\ref{fig:convergence_combined}; WINN, trained on MAD-filtered TGLF data (Section~\ref{sec:dataset}), has learned the smooth physics trend that the raw TGLF outputs imply in aggregate.

Third, a single additional flux-matching step with TGLF warm-started from WINN's converged profile drifts the profile by approximately 10--20\% in core $n_e$ and 50\% in core $\omega_0$, confirming that TGLF does not accept WINN's solution as its own flux-matched state.

We observed the same qualitative behavior when repeating this diagnostic with \base~in place of \ours: TGLF fluxes evaluated at the \base-converged profile are likewise highly oscillatory, and TGLF warm-started from the \base-converged state drifts by a comparable amount. We show only the \ours~panels for brevity. The relevant gap is therefore surrogate-vs-solver rather than surrogate-vs-surrogate: TGLF's residual surface is too noisy for Anderson-accelerated convergence regardless of which neural network replaces the turbulent module.

\begin{figure}[!h]
    \centering
    \includegraphics[width=1.0\textwidth]{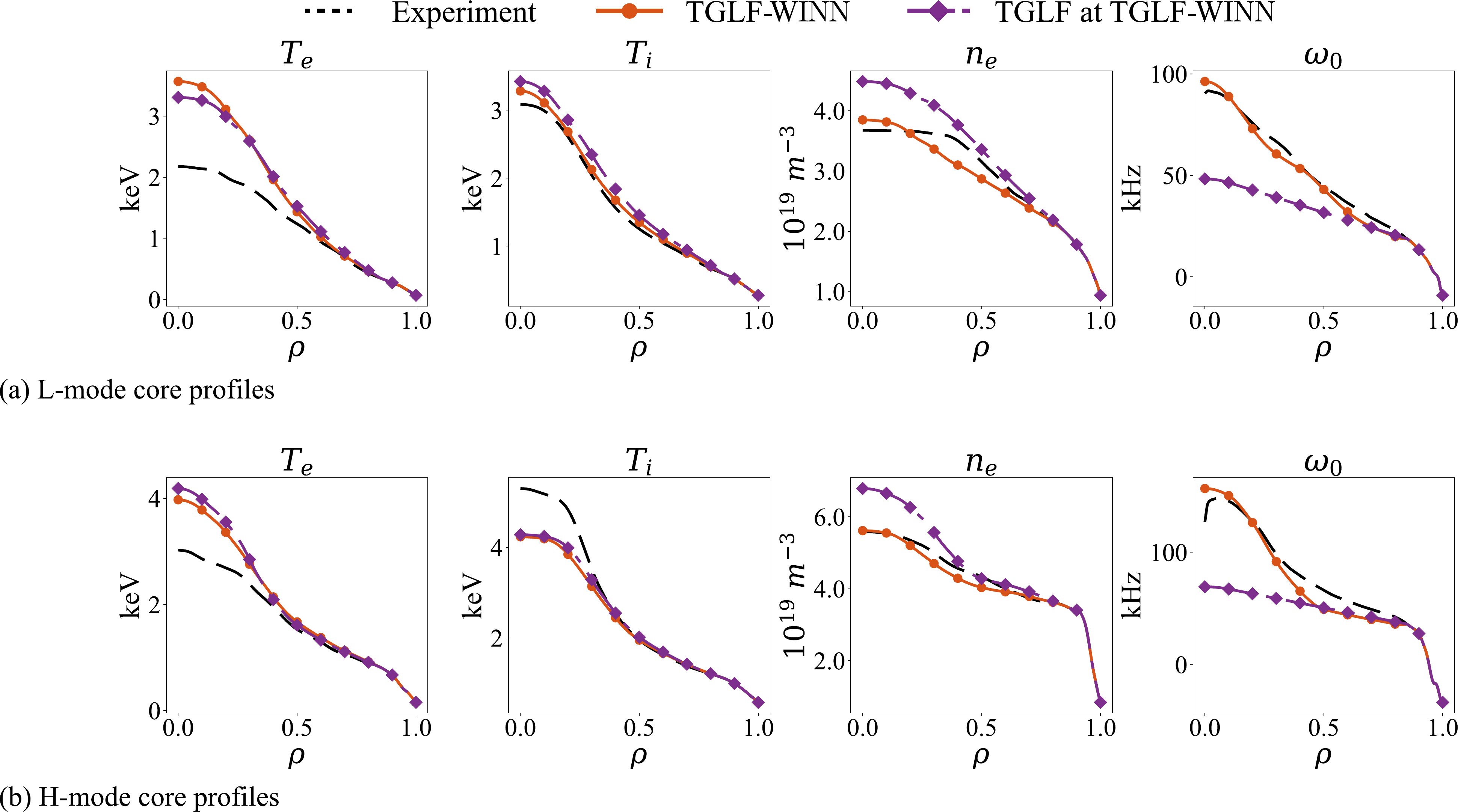}
    \caption{\revision{\textbf{Profile drift after one warm-started TGLF flux-matching step, L-mode (top) and H-mode (bottom).}} Starting from WINN's converged profile, one additional TGLF step drifts core $\omega_0$ by $\sim\!50\%$ and core $n_e$ by $\sim\!15\%$ in L-mode, and drifts core $\omega_0$ by $\sim\!50\%$ and core $n_e$ by $\sim\!20\%$ in H-mode; $T_e$ and $T_i$ remain close to WINN's profile in both cases. The drift pattern is dominated by the channels where TGLF's raw flux prediction is noisiest (cf. Figures~\ref{fig:tglf_at_nn_L} and \ref{fig:tglf_at_nn_H}).}
    \label{fig:tglf_at_nn_profiles}
\end{figure}

\section{Per-Branch Parity Analysis}
\label{sec:appendix_per_branch_parity}

\begin{figure}[!p]
    \centering
    \includegraphics[width=\textwidth,height=0.9\textheight,keepaspectratio]{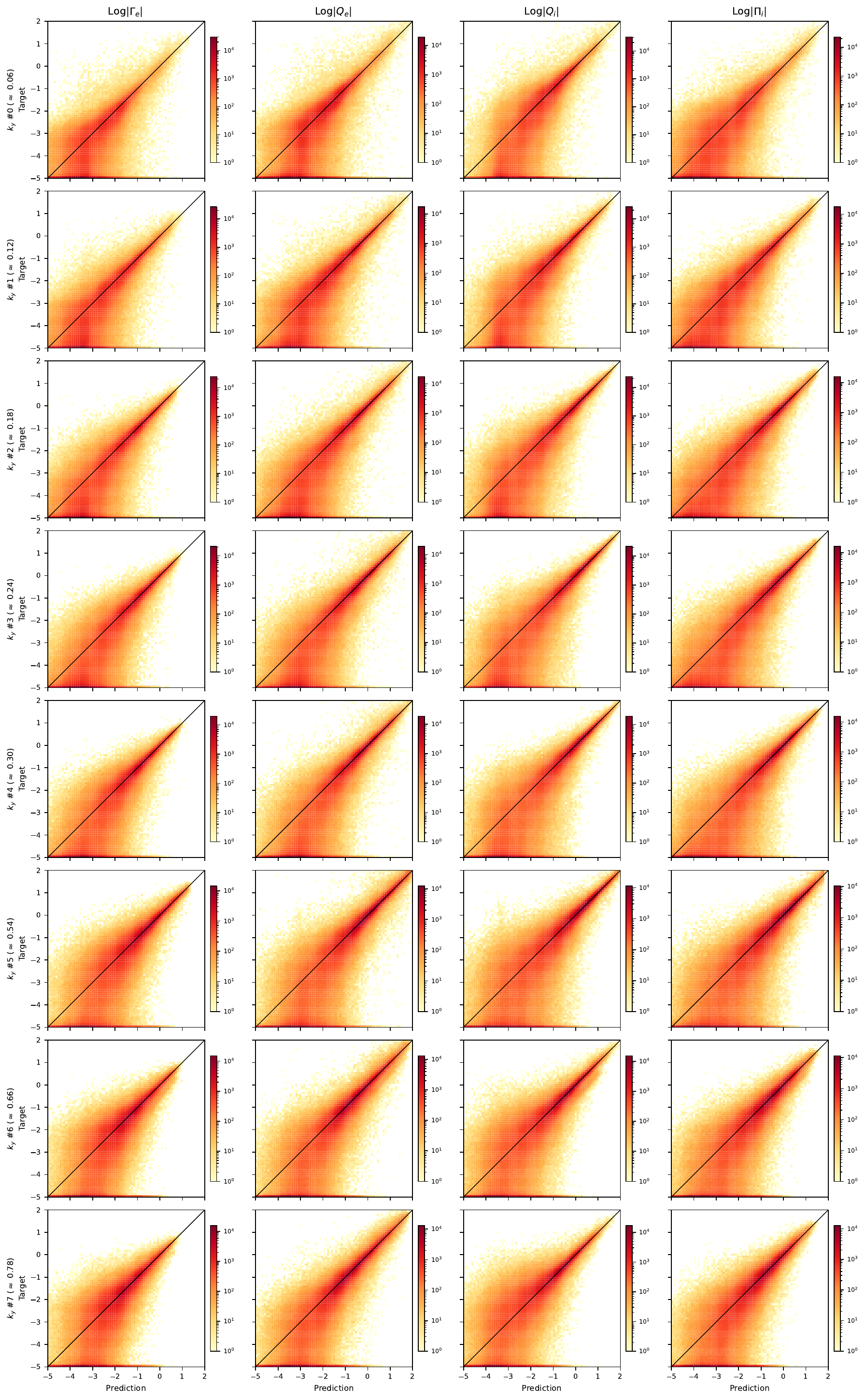}
    \caption{\textbf{Per-branch parity, low $k_y$ (branches 0--7; $k_y \rho_s \in [0.06, 0.78]$; ion-scale, ITG range).} $\log_{10}|\text{prediction}|$ vs. $\log_{10}|\text{target}|$ for each $(k_y, \text{channel})$ pair on the full offline test set. Columns: $\Gamma_e$, $Q_e$, $Q_i$, $\Pi_i$. Rows: $k_y$ branch index with its approximate dataset-mean value. Log-scaled color shows sample density; black diagonal marks exact agreement.}
    \label{fig:per_branch_parity_low}
\end{figure}

\begin{figure}[!p]
    \centering
    \includegraphics[width=\textwidth,height=0.9\textheight,keepaspectratio]{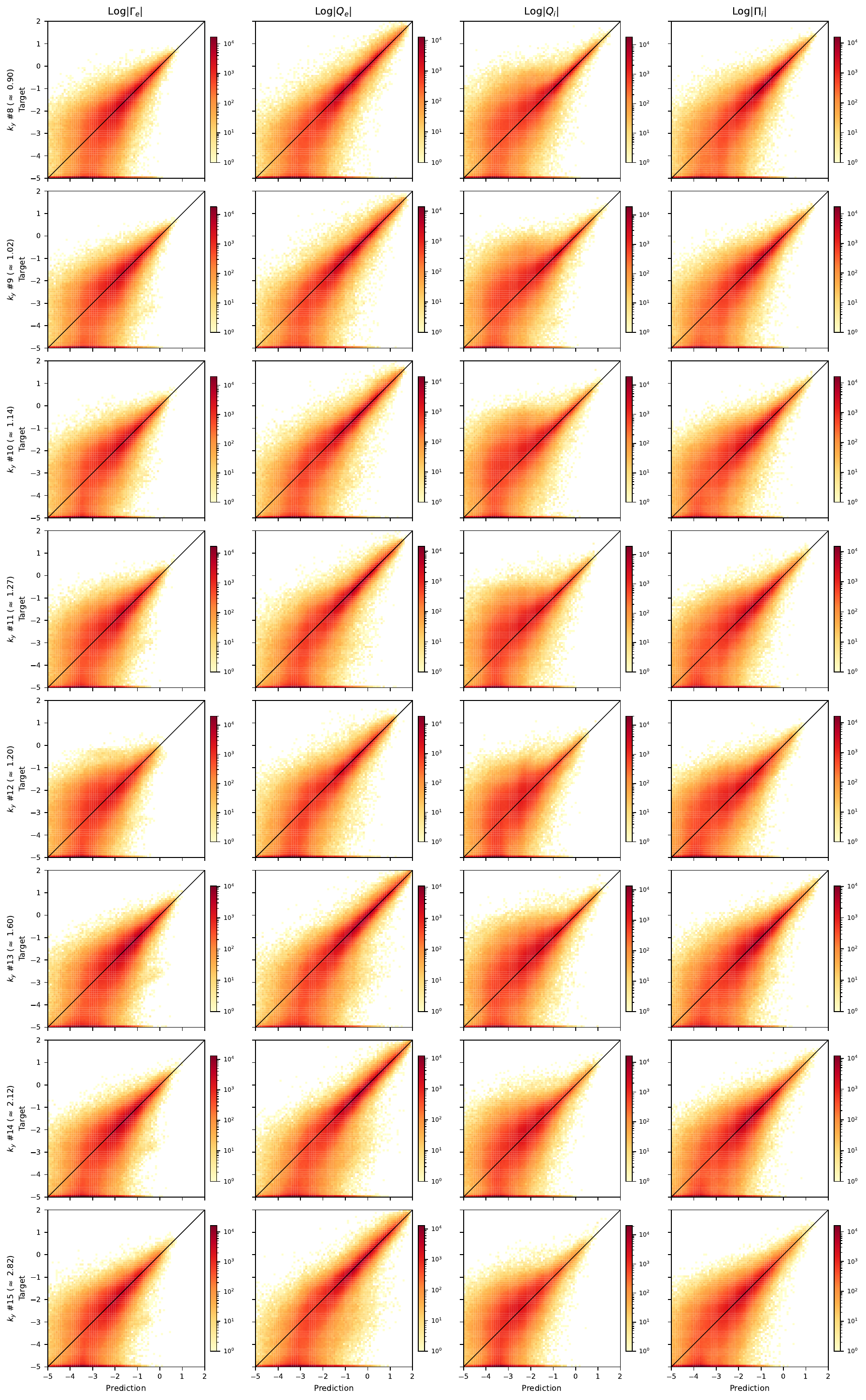}
    \caption{\textbf{Per-branch parity, mid $k_y$ (branches 8--15; $k_y \rho_s \in [0.90, 2.82]$; sub-ion-scale, TEM range).} Same layout as Figure~\ref{fig:per_branch_parity_low}.}
    \label{fig:per_branch_parity_mid}
\end{figure}

\begin{figure}[!p]
    \centering
    \includegraphics[width=\textwidth,height=0.9\textheight,keepaspectratio]{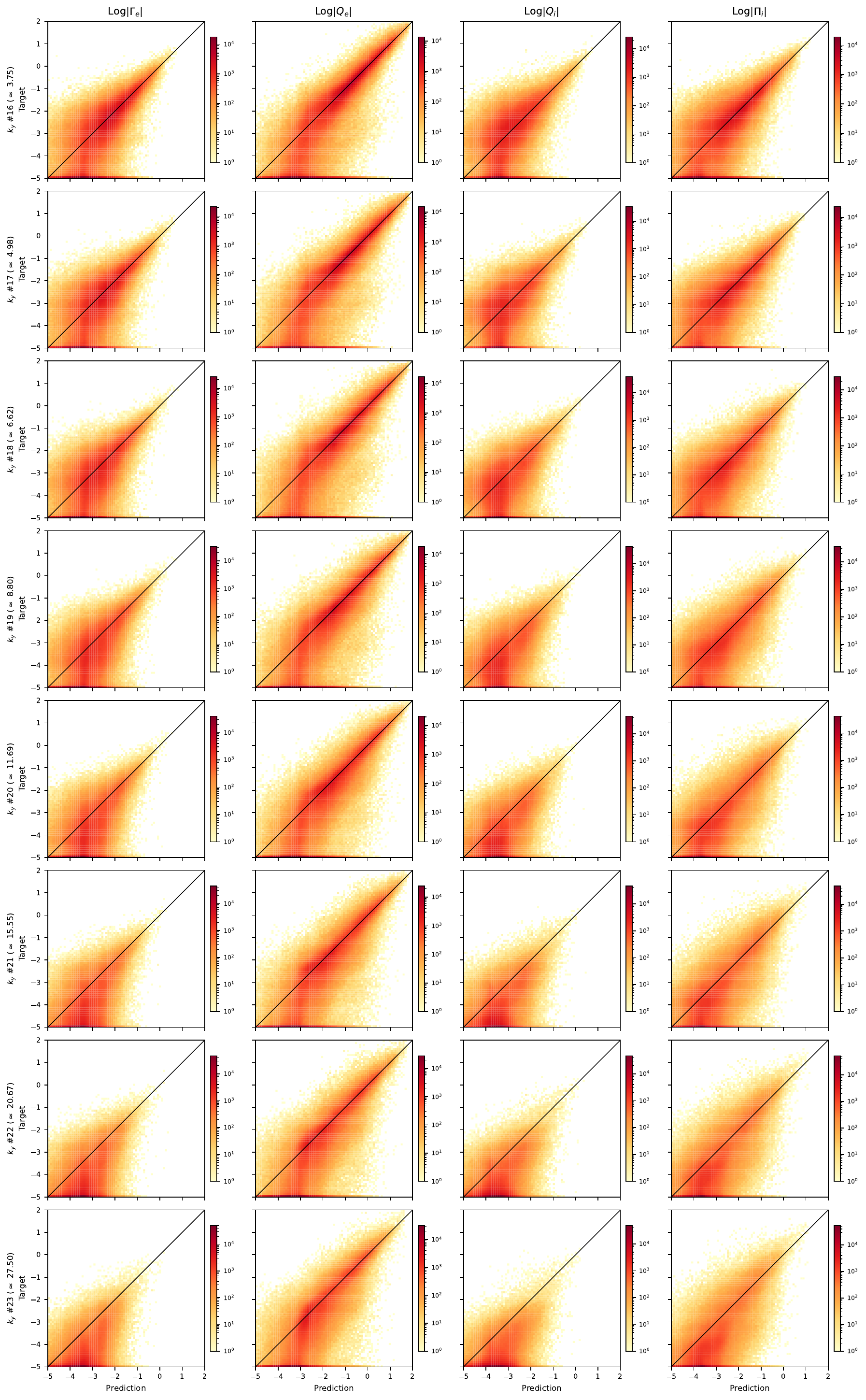}
    \caption{\textbf{Per-branch parity, high $k_y$ (branches 16--23; $k_y \rho_s \in [3.75, 27.5]$; electron-scale, TEM--ETG transition and ETG range).} Same layout as Figure~\ref{fig:per_branch_parity_low}.}
    \label{fig:per_branch_parity_high}
\end{figure}

This appendix supports the per-branch classification discussion in Section~\ref{sec:res:overall}. For each of the 24 $k_y$ branches and each of the four flux channels $(\Gamma_e, Q_e, Q_i, \Pi_i)$, we show a 2-D histogram of $\log_{10}|\text{prediction}|$ versus $\log_{10}|\text{target}|$ on the full offline test set ($N \approx 9\!\times\!10^5$ samples) using the final \ours~checkpoint. Within each panel, a black diagonal marks exact agreement. Agreement is tight along the diagonal across all low- and mid-$k_y$ branches (Figures~\ref{fig:per_branch_parity_low}, \ref{fig:per_branch_parity_mid}), and remains visibly present at high $k_y$ (Figure~\ref{fig:per_branch_parity_high}) with broader scatter for $\Gamma_e$ and $Q_i$ in the highest-$k_y$ branches where the underlying TGLF fluxes are small for a large fraction of samples.

\section{Implementation Details}
\label{sec:appendix:implementation}

This appendix reports the training configuration and computational resources used for the final \ours~run on the filtered \MaMiPert~dataset (Table~\ref{tab:ablation_results}, bold row, average RMSLE $5.83\times 10^{-2}$). Values for the \base~baseline runs use the same optimizer, schedule, batch size, and hardware unless otherwise noted.

\paragraph{Training Configuration.} We use the AdamW optimizer with $\beta_1 = 0.9$, $\beta_2 = 0.999$ (PyTorch defaults), and weight decay $10^{-4}$. The learning rate follows a cosine-decay schedule with linear warmup: peak learning rate $10^{-3}$, warmup over $17{,}500$ steps ($\tfrac{1}{10}$ of total training steps), cosine decay to end learning rate $10^{-6}$ over $175{,}000$ steps. Training runs for 200 epochs at batch size 4096 (approximately 875 steps/epoch, 175{,}000 total steps, matching the cosine decay horizon). The combined loss weight is $\lambda = 1$ in $\mathcal{L} = \mathcal{L}_f + \lambda\mathcal{L}_s$. No early stopping is used. The data split is 80\% training / 20\% test, with random seed 42.

\paragraph{Computational Resources.} All training and inference benchmarks were performed on a workstation with 1$\times$ NVIDIA GeForce RTX 4090 (24~GB VRAM), 2$\times$ AMD EPYC 7282 16-core CPUs (32 cores / 64 threads), and 129~GB of system RAM. A single final \ours~run on the filtered \MaMiPert~dataset takes approximately 20~hours of wall-clock time (19~hrs 57~min). Batched inference throughput is 0.41~$\mu$s/sample for \base~and 0.44~$\mu$s/sample for \ours~at batch size 4096; the $\sim$7\% inference-time overhead of \ours~reflects the per-wavenumber decomposition step. These inference timings are what enable the 45$\times$ speedup over TGLF reported in Section~\ref{sec:res:fluxmatching}.

\section{Bayesian Active Learning Computational Analysis}
\label{sec:appendix:bal_timing}

In this section, we provide detailed timing analysis for the Bayesian Active Learning (BAL) process, comparing the computational overhead of Expected Information Gain (EIG) acquisition against random sampling baseline.

\paragraph{Acquisition Function Overhead.}
The EIG acquisition function (Equation~\ref{eq:bal:eig}) requires additional computation at each iteration compared to random sampling. Table~\ref{tab:bal_timing} reports the wall-clock time breakdown for key components during active learning iterations.

\begin{table}[!h]
    \centering
    \caption{\textbf{Computational overhead comparison for active learning strategies.} Wall-clock time reported in seconds per BAL iteration, averaged over the first two BAL iterations. Uncertainty estimation uses 16 MC-Dropout forward passes per candidate. Candidate-proposal cost is similar for both strategies because both draw from per-radius Gaussian priors; EIG adds MC-Dropout and acquisition-score computation (including the trial-training step for the expected information gain).}
    \label{tab:bal_timing}
    \begin{tabular}{lcc}
        \toprule
        \textbf{Component}                                 & \textbf{Random Sampling (sec)} & \textbf{EIG Acquisition (sec)} \\
        \midrule
        Candidate proposal                                 & 139.18                         & 150.19                         \\
        Uncertainty estimation (MC Dropout, 16 passes)     & --                             & 0.87                           \\
        Acquisition score (incl. trial-training + EI eval) & --                             & 283.73                         \\
        Model retraining                                   & 167.76                         & 152.93                         \\
        \midrule
        \textbf{Total per iteration}                       & 306.95                         & 587.71                         \\
        \bottomrule
    \end{tabular}
\end{table}

\paragraph{Overhead vs. Data Reduction Trade-off.}
While EIG acquisition incurs additional computational cost compared to random sampling, this overhead is justified by the significant reduction in required training data (25\% of full dataset vs. 100\%). Furthermore, when extending this approach to higher-fidelity simulations such as CGYRO~\cite{candy2016high,Neiser23}, where a single evaluation takes hours on supercomputers, the acquisition overhead becomes negligible compared to the cost saved by requiring fewer simulation evaluations. For TGLF, where evaluations are relatively fast (seconds), the benefit is less pronounced but still demonstrates the viability of the BAL framework for future high-fidelity applications.

\section{Stored Thermal Energy Sanity Check against IPB98(y,2)}
\label{sec:appendix:wth_lumped}

On the extended validation set of 3000 H-mode and 1500 L-mode DIII-D profiles~\cite{abbate2024large}, we perform a lumped confinement sanity check of both NN surrogates against the IPB98(y,2) empirical scaling law. We define the stored thermal energy as the volume integral of the total thermal pressure,
\begin{equation}
    W_{\rm th} = \tfrac{3}{2} \int_V \left( n_e T_e + \sum_s n_s T_s \right)\, dV,
    \label{eq:Wth}
\end{equation}
where the integral runs over the confined plasma volume bounded by the last closed flux surface at $\rho_{\max}$ and the sum extends over thermal ion species. Because $W_{\rm th}$ is a volume-integrated scalar, it can mask opposing profile-shape errors (over-prediction at one radius cancelled by under-prediction at another); for that reason we treat it only as a lumped sanity check, with the spatially-resolved per-radius assessment in Section~\ref{sec:res:fluxmatching} serving as the primary validation evidence.

\begin{figure}[!h]
    \centering
    \includegraphics[width=0.5\textwidth]{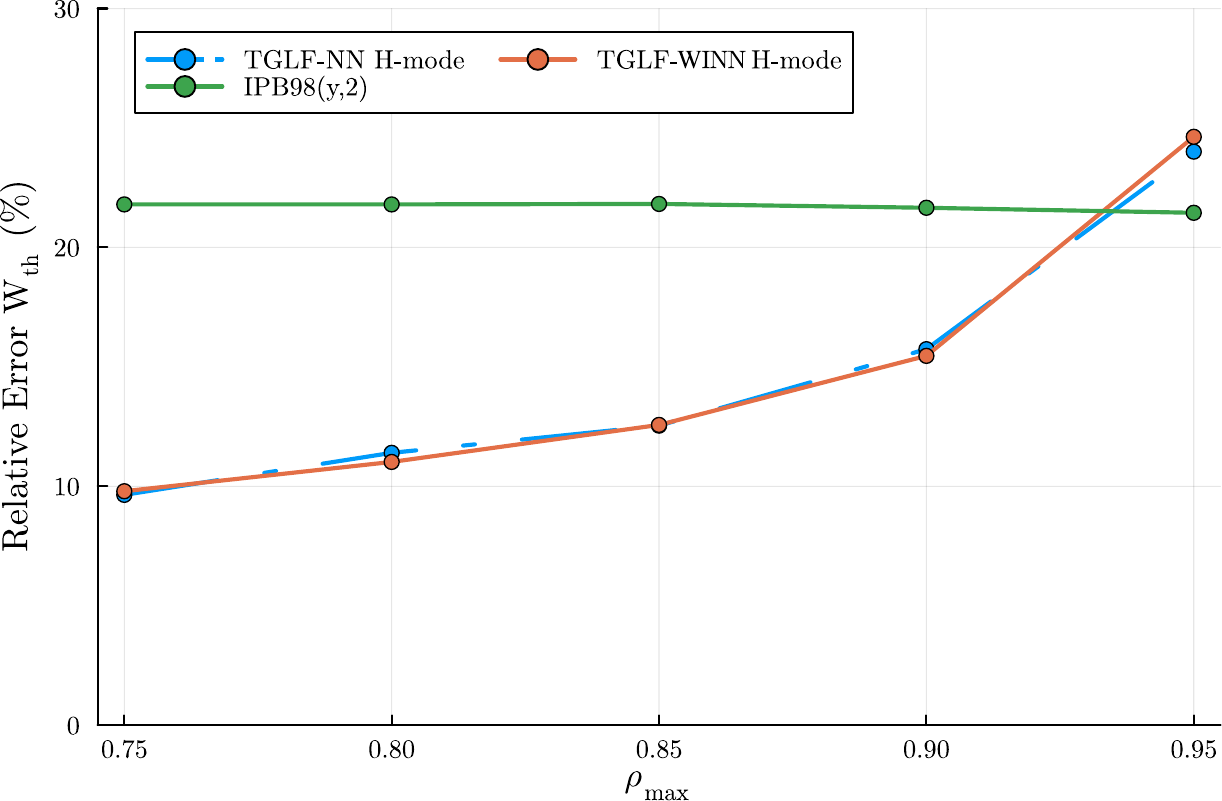}
    \caption{\revision{\textbf{Stored thermal energy ($W_{\rm th}$) predictions}.} Comparison of stored thermal energy computed from volume-integrated predicted pressure profiles for \base~and \ours~against the IPB98(y,2) scaling law across the DIII-D validation dataset. Both neural surrogate models achieve 10 percentage point higher accuracy than the IPB98(y,2) scaling law at typical $\rho_{\mathrm{max}}$ of 0.8.}
    \label{fig:wth_validation}
\end{figure}

Both neural surrogates substantially outperform the IPB98(y,2) scaling law on this benchmark (Figure~\ref{fig:wth_validation}), achieving roughly 10 percentage points higher accuracy at the typical $\rho_{\max} = 0.8$ boundary.

\section{Aggregated Validation across the Large DIII-D Dataset}
\label{sec:appendix:aggregated_validation}

\begin{figure}[!h]
    \centering
    \begin{subfigure}[b]{0.48\textwidth}
        \centering
        \includegraphics[width=\textwidth]{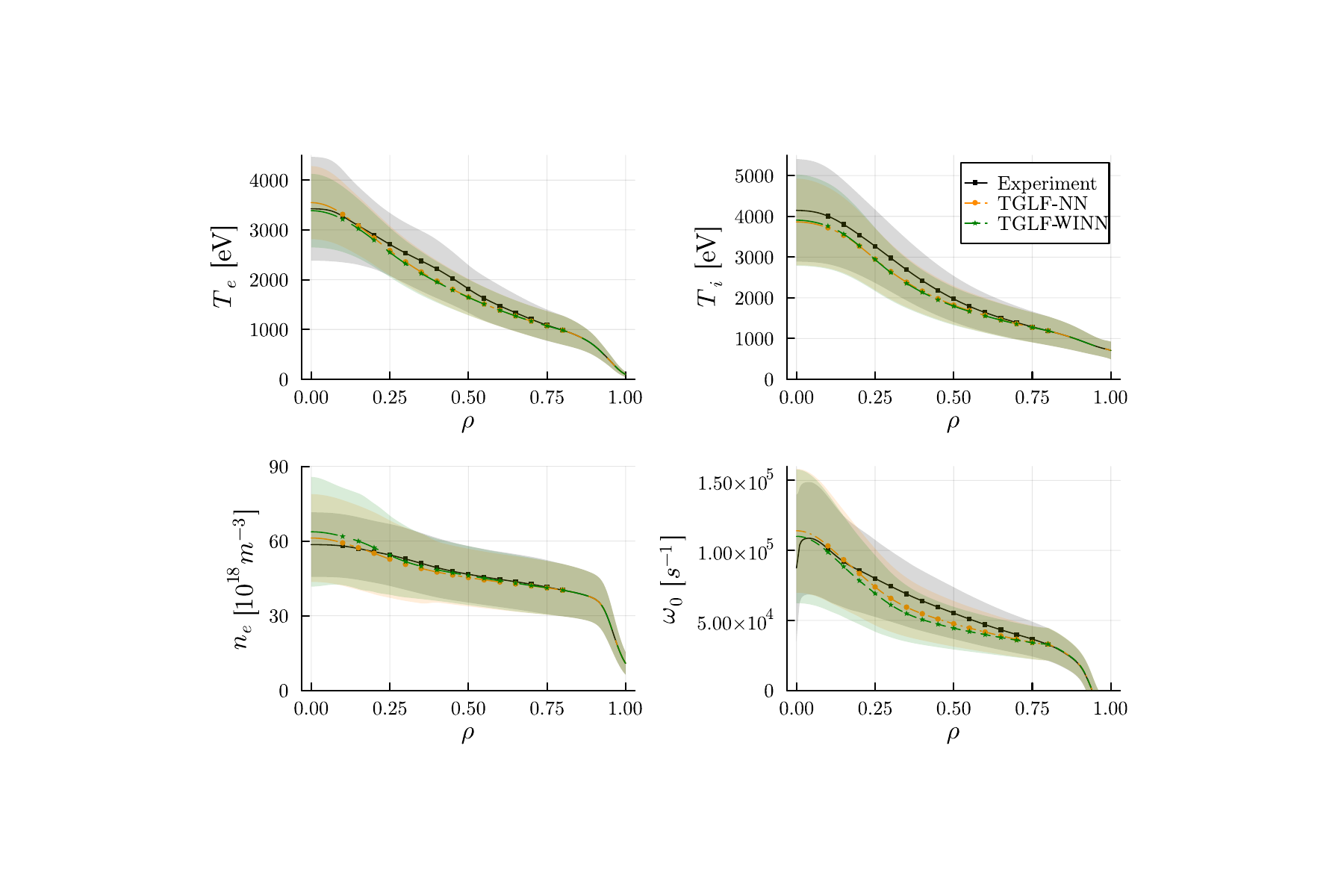}
        \caption{Profiles for 3000 DIII-D H-mode plasmas.}
        \label{fig:comp:H-mode}
    \end{subfigure}
    \hfill
    \begin{subfigure}[b]{0.48\textwidth}
        \centering
        \includegraphics[width=\textwidth]{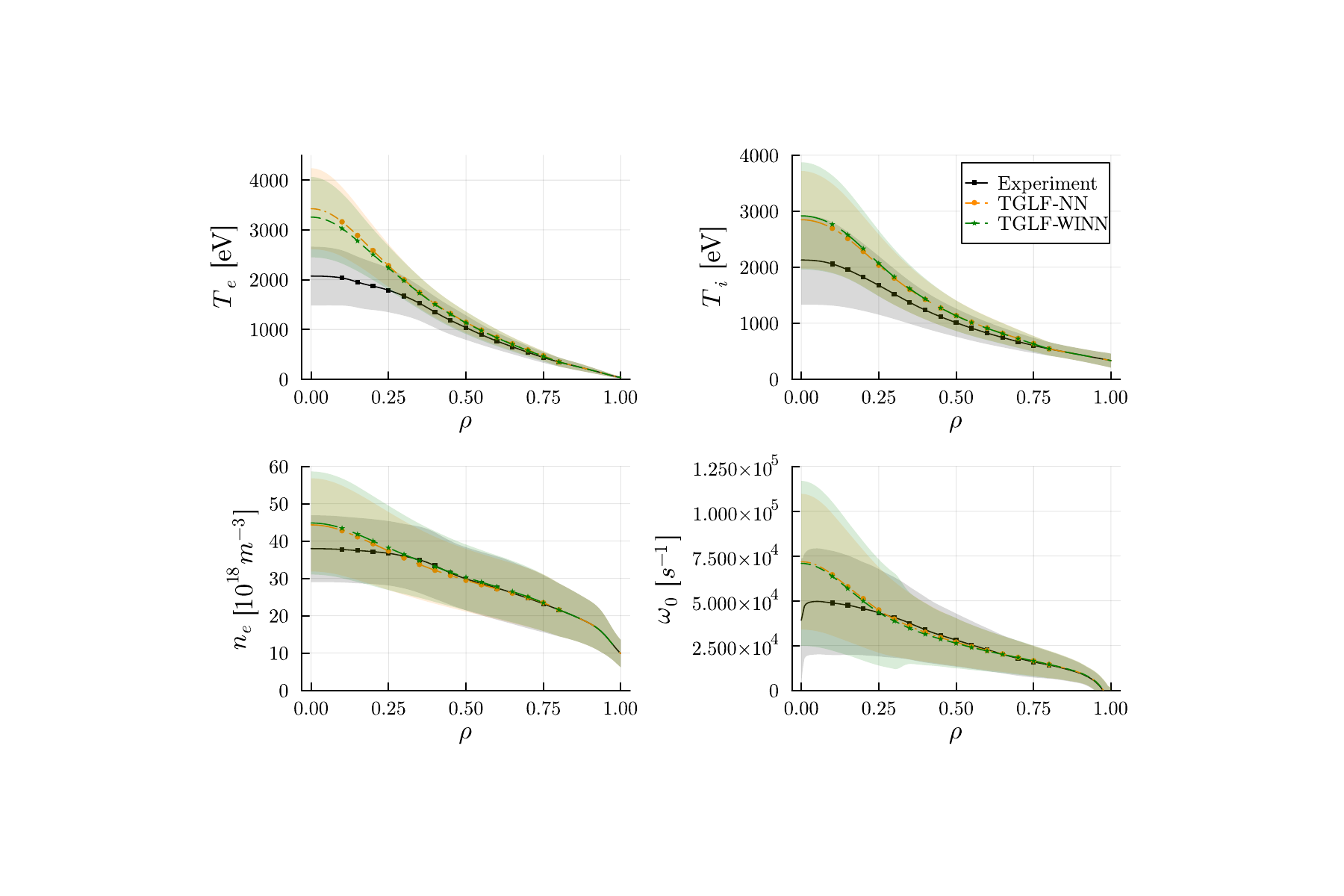}
        \caption{Profiles for 1500 DIII-D L-mode plasmas.}
        \label{fig:comp:L-mode}
    \end{subfigure}
    \caption{\revision{\textbf{Comparison of measured profiles to predictions by \base~and \ours:}} (a) 3000 H-modes and (b) 1500 L-modes with $\rho_{\rm max}=0.80$ (\emph{solid line: mean}, \emph{shaded: $\pm 1$ standard deviation}).}
    \label{fig:comparison:profile}
\end{figure}

This appendix complements the per-radius relative-error view in Section~\ref{sec:res:fluxmatching} (Figure~\ref{fig:rhomax_comp}) with an aggregated mean-and-standard-deviation comparison of predicted profiles against experimental measurements across the full 3000-H-mode and 1500-L-mode DIII-D validation dataset. We retain this distributional view because, while individual-discharge and per-radius comparisons (Figures~\ref{fig:profiles_combined} and~\ref{fig:rhomax_comp}) are the primary spatially-resolved assessments, the aggregated mean$\pm$std summary across thousands of discharges is a standard practice for large-database surrogate validation~\cite{abbate2024large} and provides a concise screen for systematic trends that would be impractical to inspect discharge-by-discharge. Both \base~and \ours~show comparable mean performance, with small systematic underprediction of temperature and rotation in H-modes and overprediction in L-mode core regions ($\rho<0.5$); these biases appear in both surrogates and therefore point to underlying TGLF ROM limitations rather than NN approximation errors (Figure~\ref{fig:comparison:profile}).

\section{Mean and Standard Deviation of Inputs Across Radial Locations}
\label{sec:appendix:RP}

\begin{figure}[!h]
    \centering
    \begin{minipage}[c][0.9\textheight][c]{1.0\textwidth}
        \centering
        \includegraphics[width=1.0\textwidth]{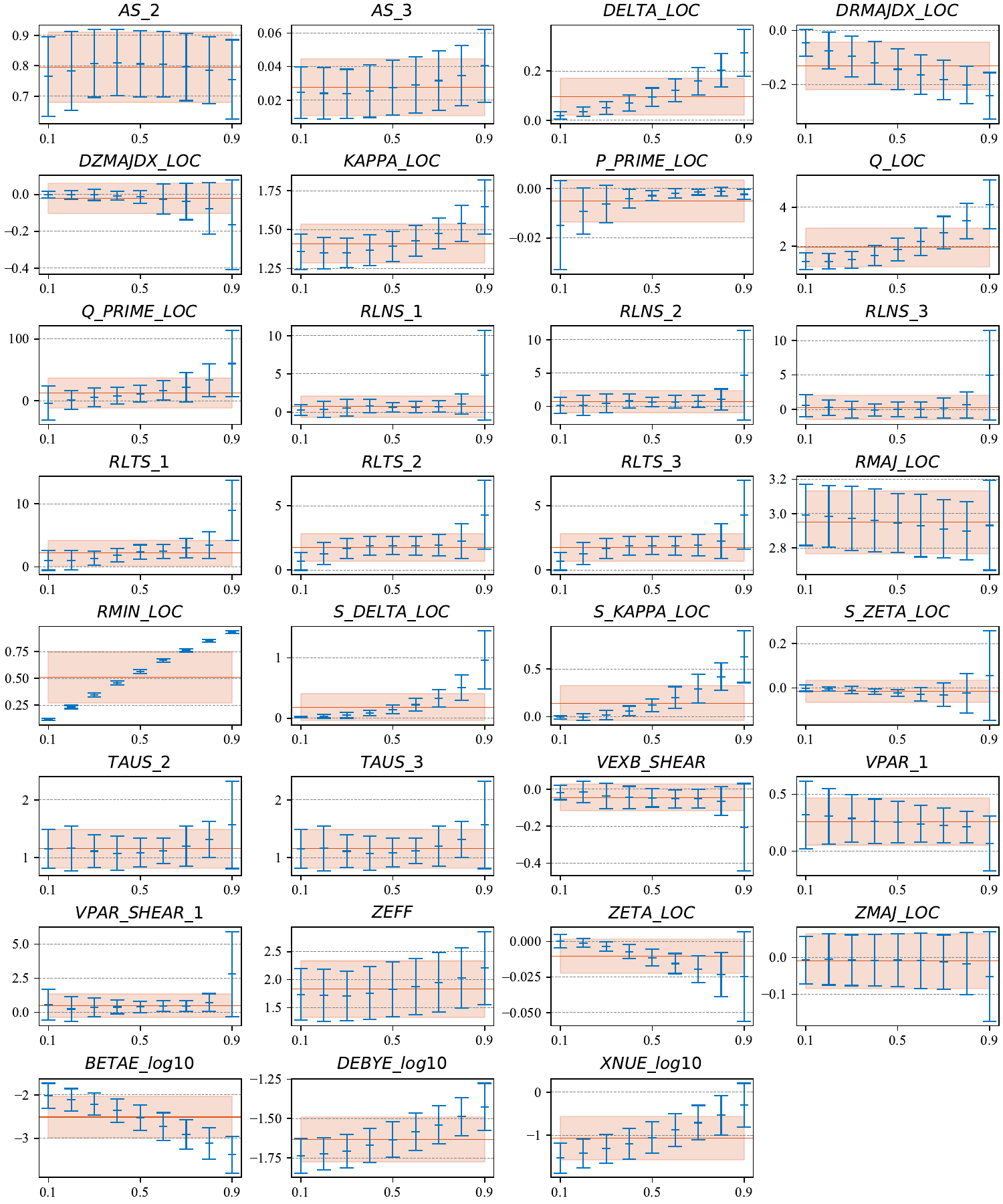}
    \end{minipage}
    \caption{Radial profiles of mean values and standard deviations for input parameters. Each subplot represents a distinct input parameter, with the x-axis showing the normalized radial coordinate $\rho$ and the y-axis displaying the parameter's value. Blue bars indicate mean values, while orange bars represent standard deviations at each radial location.}
    \label{fig:mstd_accross_rho}
\end{figure}

To obtain the statistics of input parameters for each radial location, we parse the metadata of the shot data and group each shot into its corresponding radial location bracket. After parsing the entire dataset, we compute the mean and standard deviation for each radial bracket. Figure~\ref{fig:mstd_accross_rho} plots these statistics for each input variable against the radial profile.

\end{document}